\documentclass[sigconf]{acmart}

\copyrightyear{2026}
\acmYear{2026}
\setcopyright{cc}
\setcctype{by}
\acmConference[WWW '26]{Proceedings of the ACM Web Conference 2026}{April 13--17, 2026}{Dubai, United Arab Emirates}
\acmBooktitle{Proceedings of the ACM Web Conference 2026 (WWW '26), April 13--17, 2026, Dubai, United Arab Emirates}
\acmPrice{}
\acmDOI{10.1145/3774904.3792083}
\acmISBN{979-8-4007-2307-0/2026/04}

\usepackage{amsfonts}
\usepackage{algorithmic}
\usepackage{textcomp}
\usepackage{xcolor}
\usepackage{fontawesome}
\usepackage{listings}

\usepackage{amssymb}
\usepackage{multirow}
\usepackage{bbding} 
\usepackage{hyperref}
\usepackage{color}
\usepackage{cleveref}
\usepackage{xspace}
\usepackage{pifont} 
\usepackage{amsmath}
\usepackage{amsfonts}
\usepackage{graphicx}
\usepackage[table]{xcolor}
\usepackage{balance}
\usepackage{subfigure}

\newcommand{\tool}{{\textsc{IntelGuard}}}

\lstdefinestyle{tablecode}{
    basicstyle=\fontsize{7}{8}\selectfont\ttfamily,
    breaklines=true,
    breakatwhitespace=false,
    xleftmargin=0pt,
    framexleftmargin=0pt,
    frame=none,
    columns=flexible,
    keepspaces=true,
    showstringspaces=false,
    tabsize=2
}

\begin{document}

\title{Bridging Expert Reasoning and LLM Detection: A Knowledge-Driven Framework for Malicious Packages}

%% Authors
\author{Wenbo Guo}
\affiliation{%
  \institution{Nanyang Technological University}
  \country{Singapore}
}
\email{honywenair@gmail.com}

\author{Shiwen Song}
\affiliation{%
  \institution{Singapore Management University}
  \country{Singapore}
}
\email{swsong@smu.edu.sg}

\author{Jiaxun Guo}
\affiliation{%
  \institution{Sichuan University}
  \country{China}
}
\email{mtt88176@gmail.com}

\author{Zhengzi Xu}
\authornote{Corresponding authors.}
\affiliation{%
  \institution{Imperial Global Singapore}
  \country{Singapore}
}
\email{z.xu@imperial.ac.uk}

\author{Chengwei Liu}
\authornotemark[1]
\affiliation{%
  \institution{Nanyang Technological University}
  \country{Singapore}
}
\email{chengwei.liu@ntu.edu.sg}

\author{Haoran Ou}
\affiliation{%
  \institution{Nanyang Technological University}
  \country{Singapore}
}
\email{haoran007@e.ntu.edu.sg}

\author{Mengmeng Ge}
\affiliation{%
  \institution{Nanyang Technological University}
  \country{Singapore}
}
\email{mengmeng.ge@ntu.edu.sg}

\author{Yang Liu}
\affiliation{%
  \institution{Nanyang Technological University}
  \country{Singapore}
}
\email{yangliu@ntu.edu.sg}

\renewcommand{\shortauthors}{Wenbo Guo et al.}




\begin{abstract}

Open-source ecosystems such as NPM and PyPI are increasingly targeted by supply chain attacks, yet existing detection methods either depend on fragile handcrafted rules or data-driven features that fail to capture evolving attack semantics. We present \tool, a retrieval-augmented generation (RAG) based framework that integrates expert analytical reasoning into automated malicious package detection. \tool{} constructs a structured knowledge base from over 8,000 threat intelligence reports, linking malicious code snippets with behavioral descriptions and expert reasoning. When analyzing new packages, it retrieves semantically similar malicious examples and applies LLM-guided reasoning to assess whether code behaviors align with intended functionality. Experiments on 4,027 real-world packages show that \tool{} achieves 99\% accuracy and a 0.50\% false positive rate, while maintaining 96.5\% accuracy on obfuscated code. Deployed on PyPI.org, it discovered 54 previously unreported malicious packages, demonstrating interpretable and robust detection guided by expert knowledge.

\end{abstract}
\maketitle
\section{Introduction}
\label{sec:introduction}

%background
Open-source package repositories have become indispensable to modern software development. Platforms such as NPM (JavaScript) and PyPI (Python) provide millions of reusable libraries that streamline development. However, their openness also exposes them to security threats~\cite{cisa2025npm, fortinet2025malicious, paloalto2025npm}.
A notable case occurred on September 8, 2025, when 18 NPM packages with over 2.6 billion weekly downloads were compromised, marking one of the most severe supply chain incidents in recent memory~\cite{paloalto2025npm}.
Consequently, detecting malicious packages in open-source repositories has become critical for software supply chain security.

% 当前的检测方法可以分为
% 1. 基于规则的 这里可以说有动态或者静态分析提取的规则 这些规则往往是由专家定义 需要耗费大量的人力 而且泛化性很差
% 2. 基于学习的 主要是为了弥补预定义规则的不足 有一些自动提取特征的方法 提取之后应用在机器学习算法中 这一类方法的缺点是 data-driven的因此存在机器学习算法固有的concept drift的问题，例如模型老化的问题
% 3. 基于大模型的 弥补基于学习的方法的不足 逐渐出现了data-free的方法 试图利用大模型对程序代码的理解能力去进行更高级的检测。然而，大模型本身固有的幻觉问题，会导致其结果不可信。目前

Existing detection methods can be broadly classified into three categories: rule-based, learning-based, and large language model (LLM)-based approaches.
Rule-based methods rely on predefined expert-crafted rules to identify suspicious patterns derived from static analysis ~\cite{ossgadget2021, guarddog2022} or dynamic analysis ~\cite{duan2021measuring, packj2022}. However, constructing and maintaining these rules demands significant manual effort and domain expertise.
Learning-based methods\cite{huang2024donapi, zhang2025killing, ladisa2023feasibility}  attempt to automatically extracting features from packages to train machine learning classifiers that distinguish malicious from benign packages. 
However, they are data-driven and thus suffer from the concept drift problem~\cite{lu2018learning}, as threat patterns evolve over time, the learned models become outdated, leading to degraded detection accuracy and increased false positives.
Now, LLM-based methods ~\cite{wang2025malpacdetector,gobbi2024genie,yu2024maltracker} represent a paradigm shift by leveraging large language models' ability to  analyze program behavior at an abstraction level beyond syntactic patterns without explicit feature engineering. Despite their potential, these approaches face two critical limitations. First, they suffer from the hallucination problem inherent to large models, resulting in unreliable or inconsistent predictions. Second, current LLM-based detectors lack access to systematic expert reasoning and contextual threat intelligence, limiting their ability to analyze sophisticated or context-dependent malicious packages.

To address the limitations of existing detection methods, we analyze how human security experts identify evasion-oriented malicious packages. 
Analysts at firms such as ReversingLabs and Trend Micro employ multi-layered analytical frameworks to assess the \textit{consistency between observed code behaviors and a package’s intended functionality and reason about whether the implementation aligns with its expected purpose} \textbf{(Challenge 1)}. For instance, a 
cryptocurrency wallet library exfiltrating token data to external chat 
services breaches the principle that cryptographic operations should remain 
self-contained.
%; installation scripts that modify locally installed applications 
%contradict the isolation boundaries enforced by package managers; and network 
%communication during package initialization deviates from the expectation that 
%setup routines perform only local configuration. 
However, consistency assessment addresses only the detection aspect.
A complementary challenge lies in attack logic reconstruction for forensic analysis to \textit{understand why a behavior is malicious and how it fits into attacker campaigns} \textbf{(Challenge 2)}.
Expert analysts achieve this by situating behaviors within attack lifecycle models (e.g., recognizing installation-phase exfiltration as pre-runtime credential theft) and correlating technical indicators with threat intelligence about attacker infrastructure and tactics.
Finally, these analysts systematically document this reasoning in threat intelligence reports that explain the logic connecting observed actions to malicious intent and known attack methodologies. Despite its value for both detection and forensics, \textit{this knowledge remains inaccessible to automated systems, trapped in unstructured PDFs and blog posts} \textbf{(Challenge 3)}.

To bridge the gap between human analytical reasoning and automated detection, we present the first retrieval-augmented generation (RAG)-based LLM framework, named \tool, for malicious package analysis. This framework transforms expert reasoning from threat intelligence reports into structured knowledge to enable interpretable and robust zero-shot detection.
Our framework operates in two phases.
(1) In the knowledge construction phrase, we develop a multi-stage pipeline to extract both behavioral indicators and expert analytical context from unstructured threat intelligence reports, \textit{addressing Challenge 2 and Challenge 3}.
Specifically, the pipeline identifies malicious code snippets and API-level behaviors, captures the corresponding expert reasoning chains that explain why such behaviors indicate compromise, and models contextual knowledge, such as attack phases, objectives, and infrastructure, into a structured knowledge base.
This process transforms unstructured textual expertise into a machine-interpretable representation that encodes the causal and contextual relationships underlying attack logic.
(2) In the detection phrase, when analyzing a new package, the framework performs semantic retrieval by encoding its program structure and querying the expert knowledge base for semantically similar malicious fragments and their associated reasoning.
The expert reasoning provides analytical context that guides detection.
We then design an LLM-guided semantic analyzer, which integrates the expert knowledge with program analysis results to assess whether the package’s code semantics align with its intended functionality, \textit{addressing Challenge 1.}

We evaluate \tool{} on datasets comprising 2,000 PyPI packages and 2,027 npm packages.
Experimental results show that \tool{} achieves 99\% accuracy, 99.49\% precision, and 98.50\% recall, outperforming state-of-the-art baselines by 4–9 percentage points while reducing false positive rates from up to 6.5\% to 0.5\%.
Moreover, \tool{} exhibits strong robustness against code obfuscation: on obfuscated samples, it maintains 96.5\% accuracy compared to 52.82\% for direct LLM analysis with a 43.68-point improvement.
Ablation studies confirm that retrieved expert reasoning plays a critical role, as removing it reduces accuracy by 2–3 points and increases false negatives.
To assess real-world effectiveness, we deployed \tool{} to scan PyPI.org between July and September 2025, discovering 54 previously unreported malicious packages, 24 of which were later confirmed and removed by the PyPI security team.
These results demonstrate that \tool{} not only achieves state-of-the-art (SOTA) detection accuracy but also delivers tangible security impact in real-world ecosystems.

The main contributions are summarized as follows:
\begin{itemize}
  \item We present a knowledge-driven detection approach that extracts expert reasoning from threat intelligence reports and applies it through retrieval-augmented generation, transforming unstructured analytical knowledge into actionable detection capabilities.
  \item We construct a comprehensive knowledge base from 8,024 threat reports containing 4,420 malicious code contexts with behavioral descriptions and expert reasoning chains that explain security violations through functional boundaries, execution contexts, and attack pattern principles.
  \item Evaluation on 4,087 packages demonstrates 99.0\% accuracy with 0.50\% false positives, outperforming rule-based, machine learning, and LLM-based baselines, while maintaining 96.50\% accuracy on obfuscated code where direct LLM analysis achieves only 52.82\%.
\end{itemize}

\section{Related Work}
\label{sec:relatedwork}

\textbf{(1) Malicious Package Detection.}
Package registries face persistent threats from typosquatting, dependency confusion, and supply chain poisoning~\cite{gu2023investigating,zahan2022weak}.
Attackers exploit installation scripts, obfuscate code, and manipulate metadata to appear legitimate~\cite{guo2023empirical,ferreira2021containing,vu2023bad}.
Early detection tools rely on static analysis or sandbox execution~\cite{ossgadget2021,guarddog2022,packj2022,bandit4mal2023,scalco2022feasibility,zahan2025leveraging}, yet attackers can easily evade them through encoding variations or conditional execution that detects analysis environments~\cite{zheng2024towards,shariffdeen2025detecting}.
To overcome these limitations, learning–based approaches model attack patterns as behavioral sequences. For instance, DONAPI~\cite{huang2024donapi} extracts API call sequences through dynamic instrumentation, Cerebro~\cite{zhang2025killing} organizes features by execution likelihood and fine-tunes BERT, and SpiderScan~\cite{huang2024spiderscan} constructs behavior graphs across sensitive APIs.
Recent systems further integrate multiple signals by combining code features with metadata~\cite{sun20241+,halder2024malicious}, applying code clustering~\cite{liang2023needle}, and extending to multilingual frameworks~\cite{li2023malwukong}.
However, these approaches remain largely data-driven, relying on large volumes of samples to extract low-level statistical features. Such features are often tightly coupled with specific datasets or time periods, and thus quickly become obsolete as attack techniques evolve.
More recently, LLM-based detectors have been developed to emulate the reasoning process of human security experts~\cite{wang2025malpacdetector,gobbi2024genie,yu2024maltracker}.
Despite their interpretability potential, these models still lack domain-specific knowledge and sufficient semantic context, often leading to unreliable judgments. As a result, legitimate packages with uncommon structures or behaviors are frequently misclassified as malicious~\cite{ladisa2023feasibility,sejfia2022practical}.

\noindent{\textbf{(2) Threat Intelligence Analysis.}}
Threat intelligence reports document attack campaigns and explain why 
observed behaviors indicate compromise~\cite{li2022attackg, siracusano2023time, xu2024intelex, rani2024ttpxhunter, hu2024llm}. 
Recent work extracts structured knowledge from these reports through 
knowledge graph construction. Open-CyKG~\cite{sarhan2021open} uses 
attention-based extraction to organize attack entities and relationships, 
while CTIKG~\cite{huang2024ctikg} and LLM-TIKG~\cite{hu2024llm} employ 
language models to identify entities and techniques, mapping them to 
frameworks like MITRE ATT\&CK. PackageIntel applies similar techniques 
to package ecosystems, extracting vulnerability details and attack 
patterns~\cite{guo2024packageintel}.
However, these systems extract high-level entities (i.e., attack techniques, 
compromised assets, threat behaviors) that are difficult to apply directly 
to code-level detection. A single abstract entity like "data exfiltration" 
maps to numerous concrete implementations using different APIs, encodings, 
and obfuscation techniques across varying contexts. Entity relationships 
alone provide insufficient guidance for identifying whether specific code 
exhibits malicious behavior.
\vspace{-10pt}

% Our work extracts expert reasoning that bridges this gap. We capture how 
% analysts connect threat concepts to concrete code: why implementation 
% choices violate expected behavior, how API usage patterns indicate 
% compromise, and what contextual factors distinguish malicious from 
% legitimate code. By organizing these reasoning chains with code examples 
% and applying them through retrieval-augmented generation, we enable 
% detection systems to reason about code like security experts.

\section{Background}
\label{sec:background}

\textbf{(1) Sensitive API.}
A sensitive API refers to any application programming interface that 
accesses critical system resources or performs privileged operations, 
such as file system access, network communication, process execution, 
and dynamic code execution.
Malicious packages primarily achieve their attack objectives by invoking 
these sensitive APIs.
For instance, using file system APIs like 
\texttt{fs.readFile()} to steal SSH keys from \texttt{\textasciitilde/.ssh/} 
or environment access APIs to exfiltrate credentials.
In our work, we adopt the existing sensitive APIs from prior works on malicious 
package analysis~\cite{yafei2023libscan, huang2024donapi, huang2024spiderscan}.
By focusing on these sensitive APIs, we can identify the most security-critical 
code snippets from large volumes of benign code, enabling our analysis to 
pinpoint subtle malicious behaviors that truly indicate compromise.

\noindent{\textbf{(2) Attack Workflow and Detection.}}
Software supply chain attacks follow a three-step workflow~\cite{sejfia2022practical, huang2024donapi}: 
(1) publish a malicious package to the registry; (2) deceive users 
into installing it; (3) execute the malicious payload. 
The first two 
steps are intrinsically linked: attackers employ various deception 
techniques such as typosquatting, dependency confusion, or account 
compromise to increase installation likelihood. These steps constitute 
the \textit{preparation phase}, while the payload execution represents 
the \textit{execution phase}.
Detection approaches target different phases of this workflow. 
\textit{Pre-installation detection} focuses on identifying malicious 
packages before they are widely adopted, analyzing package metadata, 
code patterns, and behavioral anomalies~\cite{wyss2022wolf, li2023malwukong}. \textit{Post-installation 
detection} monitors runtime behaviors or examines installed code for 
malicious activities~\cite{zimmermann2019small}. Our work focuses on pre-installation 
detection through static code analysis, aiming to identify malicious 
packages by reasoning about their code behaviors before they can 
compromise user systems.

\noindent{\textbf{(3) Motivation.}}
Current malicious package detectors~\cite{guo2023empirical, vu2023bad, ossgadget2021, guarddog2022} model behaviors as isolated actions, leading to high false negatives.
In reality, malicious intent arises not from individual actions but from behavioral sequences under specific conditions, which form a reasoning chain connecting observations to verdicts.
While recent work attempts to model such context through structured templates or predefined patterns~\cite{huang2024donapi}, these approaches assume a fixed contextual structure.
In practice, attack implementations vary widely in action sequences, API usage, and obfuscation. A single threat concept such as credential exfiltration can manifest through countless implementation variants~\cite{guo2023empirical}.
This diversity defeats rule-based modeling. Effective detection requires 
flexible reasoning that can interpret arbitrary contexts, similar to how 
security analysts approach novel attacks. Our key insight is that threat 
intelligence reports document how experts reason about malicious 
code in context.
By extracting these reasoning chains and applying them through RAG-based LLM, we enable context-aware detection that adapts to diverse attack patterns without manual encoding.

\section{Methodology}
\label{sec:methodology}

\begin{figure}[]
    \centering
    \includegraphics[width=0.46\textwidth]{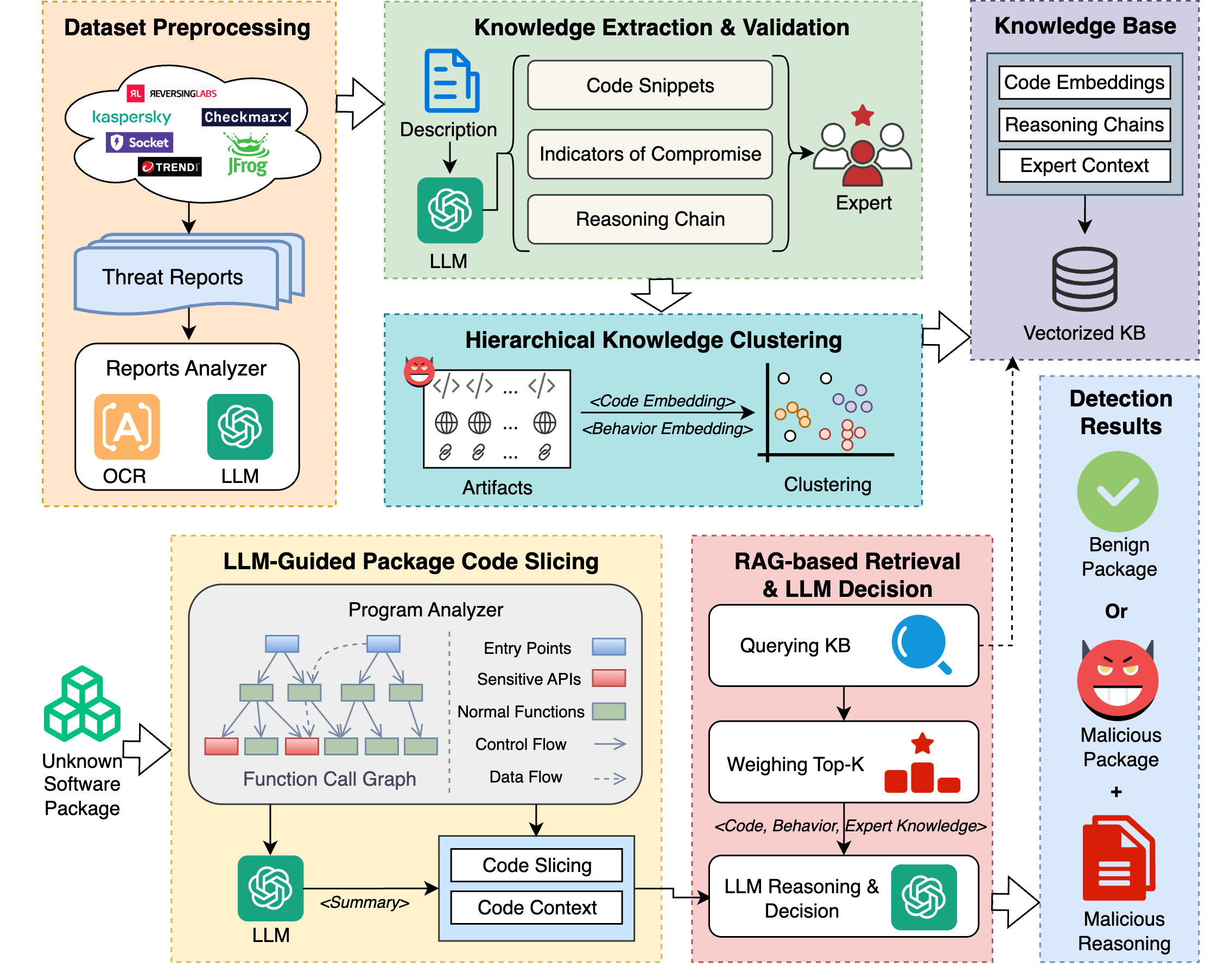}
    \caption{Framework of the \tool{}}
    \label{fig:framework}
\end{figure}

\textbf{Overview.}
We propose \tool, a knowledge-driven malicious package detection framework that integrates expert reasoning from threat intelligence reports into an automated LLM analysis pipeline.
As illustrated in Figure~\ref{fig:framework},
\tool~ operates in two major phases: \textit{knowledge construction} and \textit{malicious package detection}.

In the knowledge construction phase, we first preprocess threat intelligence reports from multiple security vendors using OCR~\cite{tesseract2025} and LLM to extract textual and visual content.
We then perform knowledge extraction and validation, identifying malicious code snippets, indicators of compromise, and expert reasoning chains that explain why specific behaviors indicate compromise \textit{(addressing Challenge 2)}.
Next, we embed each snippet using dual representations that capture both syntactic structure and behavioral semantics, followed by hierarchical knowledge clustering to organize similar behaviors.
The validated and vectorized entries are stored in a knowledge base, where each record links code fragments with their expert analytical context \textit{(addressing Challenge 3)}.

In the malicious package phase, \tool~ receives an unknown software package and applies an LLM-guided code slicing procedure to isolate functional segments surrounding sensitive API calls.
Each code slice is then used to query the knowledge base, retrieving semantically similar malicious examples and their associated expert reasoning.
Finally, a LLM performs RAG-based reasoning and decision-making, leveraging retrieved knowledge to determine whether the observed behaviors align with known attack patterns or violate functional expectations \textit{(addressing Challenge 1)}.
The system outputs both the detection result and an interpretable reasoning report explaining the decision process.

\subsection{Knowledge Construction Phrase}

This phase serves as the knowledge foundation for the subsequent malicious detection phase.
It addresses two key challenges: automated systems’ inability to access the analytical reasoning embedded in human-written threat reports, and the lack of interpretability in their detection decisions.
By systematically extracting expert reasoning chains, behavioral indicators, and contextual code knowledge from threat intelligence reports, this phase constructs a queryable knowledge base that encapsulates human analytical logic, enabling consistent and interpretable analysis in later detection.

\subsubsection{Dataset Preprocessing}
\label{sec:report_collection}

To build a comprehensive expert knowledge base for subsequent detection, we collect and preprocess threat intelligence reports from major security vendors that document expert analyses of real-world malicious packages.  
The preprocessing pipeline consists of four stages:

\noindent\textbf{(1) Identifying confirmed malicious packages.}
We query the Open Source Vulnerabilities (OSV) database~\cite{osv2025} to obtain verified malicious packages in npm and PyPI ecosystems. OSV is a widely-used open-source vulnerability database maintained by Google's Open Source Security Team. This provides a trusted seed list of known threats for subsequent analysis.

\noindent\textbf{(2) Locating relevant threat reports.}
To collect expert analyses for each malicious package, we construct search 
queries combining the package name, version, and security-related keywords 
(e.g., "malicious", "backdoor", "supply chain attack") following existing 
work~\cite{guo2024packageintel}. We submit these queries to the Google 
Search API to retrieve relevant security reports, advisories, and incident 
analyses from diverse online sources.

\noindent\textbf{(3) Extracting and reconstructing textual content.}
Security vendor reports exhibit heterogeneous HTML structures and often embed code as screenshots rather than text. To preserve complete expert analysis, we develop custom web scrapers that extract both textual content and code images while recording their spatial positions in the original layout. We then apply OCR~\cite{tesseract2025} to convert code screenshots into text and reinsert  the recognized code at its original position based on the recorded layout information. This reconstruction process ensures that code snippets remain contextually aligned with their surrounding explanations, preserving the 
semantic coherence necessary for accurate knowledge extraction.

\noindent\textbf{(4) Filtering relevant content.}
From the 8,024 pages collected across 22 prominent security sources (e.g., ReversingLabs~\cite{reversinglabs2025}, Trend Micro~\cite{trendmicro2025}, and Kaspersky~\cite{kaspersky2025}), many contain only brief vulnerability announcements or advisories without technical details, while others provide in-depth malicious package analyses.
We employ an LLM-based filter to identify pages that contain malicious code samples accompanied by expert interpretation of their behavior. 
Specifically, we prompt the LLM to assess whether each page provides actionable threat intelligence, discarding pages that contain only vulnerability metadata or high-level advisories. This filtering reduces the dataset to 4,420 pages containing substantive expert analyses suitable for knowledge extraction.

\subsubsection{Malicious Code and Reasoning Extraction}
\label{sec:knowledge_extraction}

We extract structured knowledge through two steps: extracting malicious code with expert reasoning, and validating extraction accuracy.

\noindent\textbf{Extracting structured knowledge.} 
Security experts typically compose threat reports by interweaving textual analysis with code screenshots, embedding snippets of malicious code at specific positions within their narratives.
To preserve this analytical structure, we apply OCR~\cite{tesseract2025} to extract code from embedded images and reinsert the recognized text into its original position within the web page.
This process reconstructs a unified document that retains the expert’s logical presentation.
We then feed the reconstructed text to a large language model with a predefined extraction prompt.
The model outputs structured knowledge that includes the execution context of each malicious code snippet (i.e., trigger, file location, permissions), behavioral summaries describing what the code does, expert analytical reasoning explaining why certain behaviors violate security principles, and detection indicators.
Each code snippet is treated as a basic analytical unit since it represents a concrete attack behavior with measurable context.

\noindent\textbf{Validating extraction accuracy.} 
LLMs may hallucinate non-existent reasoning or misinterpret code semantics during extraction.
To ensure the reliability and quality of the constructed knowledge base,
we introduce a human-in-the-loop validation stage.
In this process, the language model reviews both the original threat report and the extracted structured knowledge,
and a subset of the extractions is randomly audited by two security experts,
each with over five years of experience in malware analysis.
These experts evaluate whether the extracted reasoning faithfully reflects the analyst’s interpretation,
whether behavioral descriptions accurately capture code functionality,
and whether detection indicators correspond to concrete evidence mentioned in the report.
Only extractions that pass both automated and expert validation are incorporated into the knowledge base,
ensuring that stored knowledge remains grounded in authentic human analytical reasoning.

\subsubsection{Hierarchical Knowledge Clustering}
\label{sec:knowledge_clustering}

After validating the extracted knowledge, we aim to consolidate semantically related malicious behaviors to form higher-level functional representations.
Since multiple threat reports may describe the same malicious functionality using different implementations or obfuscation techniques,
clustering enables \tool~to merge heterogeneous yet functionally equivalent knowledge into unified behavioral concepts that support efficient retrieval and reasoning.

\noindent\textbf{Clustering pipeline.}
We perform a two-level hierarchical clustering process to jointly capture syntactic and semantic similarities.
(1) At the lower level, we cluster code embeddings $\mathbf{E}^{\text{code}}$ using cosine similarity to group snippets with comparable syntactic structures or lexical patterns.
We adopt HDBSCAN~\cite{mcinnes2017hdbscan} as the clustering algorithm because it automatically infers the number of clusters and is robust to noise caused by minor lexical variations or formatting differences.
We empirically set the minimum cluster size to 8 and the minimum samples parameter to 4 to balance cohesion and coverage at the code level.
(2) At the upper level, we cluster behavioral embeddings $\mathbf{E}^{\text{be}}$ derived from expert reasoning and behavioral summaries using the same HDBSCAN algorithm but with larger parameters (i.e., minimum cluster size = 40, minimum samples = 8, $\epsilon = 0.45$) to accommodate broader semantic diversity.
This hierarchical procedure abstracts away from syntactic differences while aligning functionally equivalent snippets that express the same malicious intent.

\noindent\textbf{Cluster representation.}
For each behavior-level cluster $C_j$, we compute a centroid embedding
$\mathbf{v}j = \frac{1}{|C_j|}\sum{i\in C_j}\mathbf{e}_i^{\text{be}}$
as its representative vector and select the snippet whose embedding is closest to the centroid as the canonical representative.
We further aggregate the expert reasoning chains of its member snippets using majority voting over key reasoning predicates (e.g., data exfiltration, persistence, code injection) to form a unified behavioral explanation.
This process not only reduces redundancy but also establishes a structured mapping from
\textit{implementation-level code artifacts} to \textit{behavior-level malicious intents}.
The resulting clustered knowledge is stored in the vectorized knowledge base as compact and semantically organized entries, serving as retrieval anchors for subsequent RAG-based detection.

\subsubsection{Knowledge Base Construction}
\label{sec:kb_construction}

After clustering, we construct a vectorized knowledge base to enable similarity-based retrieval during malicious package detection.

\noindent\textbf{Embedding generation.}
For each validated and clustered snippet, we compute dual embeddings that jointly encode syntactic and semantic properties. 
Given a snippet $s_i$ and its behavioral summary $b_i$, we derive:
\begin{itemize}
\item \textit{Code embedding} $\mathbf{e}_{i,\text{code}} = f_{\text{code}}(s_i)$, capturing structural and lexical characteristics of the code.  
\item \textit{Behavior embedding} $\mathbf{e}_{i,\text{behav}} = f_{\text{behav}}(b_i)$, capturing semantic intent and high-level functionality described by expert reasoning.
\end{itemize}
The dual-embedding design ensures that syntactically distinct but functionally equivalent code variants (e.g., obfuscated or reimplemented attacks) can still be aligned through their behavioral representations.

\noindent\textbf{Knowledge base schema.}
Each knowledge entry is stored as a structured tuple:
$$
\mathcal{E}_i = (s_i, c_i, b_i, r_i, d_i, \mathbf{e}_{i,\text{code}}, \mathbf{e}_{i,\text{behav}}),
$$
where $s_i$ denotes the code snippet, $c_i$ the execution context (i.e., trigger, location, permissions), $b_i$ the behavioral description, $r_i$ the expert reasoning, and $d_i$ the detection indicators.  
All entries collectively form the knowledge base $\mathcal{KB} = \{\mathcal{E}_1, \ldots, \mathcal{E}_N\}$, represented by two embedding matrices:
$$
\mathbf{E}_{\text{code}} \in \mathbb{R}^{N \times d_c}, \quad
\mathbf{E}_{\text{behav}} \in \mathbb{R}^{N \times d_b}.
$$
This representation enables efficient similarity-based retrieval of relevant malicious behaviors in subsequent detection.

\subsection{Malicious Package Detection Phrase}

This phase serves as the detection component of our framework.
It addresses the \textit{Challenge 1}, which concerns detecting inconsistencies between program semantics and intended functionality, by leveraging the expert knowledge constructed in the previous phase.
Specifically, we (1) isolate potentially malicious logic through sensitive code slicing, (2) retrieve semantically similar malicious examples and expert reasoning via RAG, and (3) perform LLM-based reasoning to determine whether the target behavior violates expected functionality.
Together, these steps enable knowledge-guided and interpretable semantic consistency analysis.

\subsubsection{Sensitive Code Slicing}
Direct static analysis of entire packages is computationally expensive and 
introduces substantial noise from benign code. We leverage a key observation: 
malicious logic inevitably invokes sensitive APIs to interact with the system 
or network~\cite{huang2024donapi, huang2024spiderscan,zheng2024towards,zhang2025killing,liang2023needle}. 
This enables us to focus analysis on code regions most likely to exhibit 
malicious behavior.

\noindent\textbf{Sensitive API identification.}
We compile a comprehensive list of sensitive APIs by systematically reviewing 
existing detection studies~\cite{huang2024donapi, huang2024spiderscan}, 
covering five categories representing common attack vectors: network operations, 
encryption operations (Base64 encoding, cryptographic functions), process 
operations (command execution, shell spawning), file operations, and system 
information operations. We treat all methods within these modules as sensitive 
APIs, which remain stable across language versions and provide broad coverage 
without frequent updates.

\noindent\textbf{Complete behavior extraction.}
To capture complete behavioral context rather than isolated API calls, we 
model a malicious behavior as a path from an \textit{entry point} (where 
execution begins) to a \textit{sensitive API} (where it manifests externally). 
Extracting this complete path requires capturing both what data the API 
operates on and under what conditions it executes. We achieve this through 
backward program slicing with two complementary techniques:

\noindent\textit{(1) Backward data-flow slicing} traces how data flows into the sensitive 
API's parameters, ensuring parameter context completeness. This captures all 
data sources, variable definitions, and transformations that influence what 
the API operates on.

\noindent\textit{(2) Backward control-flow slicing} identifies conditions governing the 
API's execution, ensuring invocation context completeness. This captures 
conditional statements, branching logic, and control dependencies that 
determine when and under what circumstances the API is called.

\noindent\textbf{Implementation.}
We construct static call graphs for each package using PyCG~\cite{PyCG2023} 
for Python (through static type inference and control flow analysis) and 
Jelly~\cite{jelly2025} for JavaScript (handling dynamic features such as 
callbacks and prototype chains). Starting from each sensitive API call site 
as the slicing criterion, we perform backward dependency tracking to identify 
all statements that may influence the API invocation. The resulting code slice 
constitutes a self-contained functional unit spanning from entry point to 
sensitive API, preserving the complete data transformations and control 
conditions necessary for understanding the behavior.

\subsubsection{RAG-based Similarity Retrieval}

Given a code slice $q$ extracted from a target package, we retrieve the most relevant malicious examples from the knowledge base to provide expert analytical context for detection. 
We first summarize the slice’s behavior $b_q$ (e.g., \textit{reads configuration files and sends data to an external server}), and then compute two embeddings using the same encoders from the knowledge construction phase: 
a code embedding $\mathbf{e}_q^{\text{code}}$, and 
a behavior embedding $\mathbf{e}_q^{\text{behav}}$ capturing semantic intent.

% \noindent\textbf{Similarity computation.}
For each knowledge entry $\mathcal{E}_i \in \mathcal{KB}$, we compute two cosine similarities:
\[
\text{sim}_{\text{code}}(q,i)=\cos(\mathbf{e}_q^{\text{code}},\mathbf{e}_i^{\text{code}}), 
\]

\[
\text{sim}_{\text{behav}}(q,i)=\cos(\mathbf{e}_q^{\text{behav}},\mathbf{e}_i^{\text{behav}}).
\]

The overall similarity is aggregated as:
\[
\text{sim}_{\text{total}}(q,i)=\alpha \cdot \text{sim}_{\text{code}}(q,i)+\beta \cdot \text{sim}_{\text{behav}}(q,i),
\]
where $\alpha=0.5$ and $\beta=0.5$ balance syntactic and semantic matching. 
This design prioritizes behavioral similarity, which remains robust under code obfuscation or minor lexical variations.

\noindent\textbf{Retrieval output.}
The top-$k$ entries $\{\mathcal{E}_{i_1},\ldots,\mathcal{E}_{i_k}\}$ with the highest total similarity scores are retrieved as contextual exemplars. 
Each entry provides its code snippet $(s_{i_j})$, behavioral description $(b_{i_j})$, and expert reasoning chain $(r_{i_j})$ that explains why the historical case was deemed malicious. 
These retrieved exemplars form the analytical grounding for subsequent reasoning.

\subsubsection{Maliciousness Determination}

We determine package maliciousness by analyzing each sensitive code slice independently through LLM-based reasoning augmented with expert knowledge. 
For each slice $q$ with its behavioral summary $b_q$, we construct a detection prompt containing three components: 
(1) the target slice and its behavioral description, 
(2) the retrieved malicious examples $(s_{i_1}, \ldots, s_{i_k})$ and their behavioral summaries $(b_{i_1}, \ldots, b_{i_k})$, and 
(3) the corresponding expert reasoning chains $(r_{i_1}, \ldots, r_{i_k})$ that articulate why these examples violate security principles.

\noindent\textbf{LLM reasoning.}
The LLM is instructed to assess whether the target slice exhibits similar security violations by comparing its behavior with the retrieved cases and applying the expert analytical principles. 
It outputs a binary label $\ell_q \in \{\text{malicious}, \text{benign}\}$ and an accompanying explanation $e_q$ that specifies which security principles are violated and how the observed behavior aligns with known attack patterns.

\noindent\textbf{Package-level aggregation.}
A package $p$ is classified as malicious if any of its slices receives a malicious verdict:
\[
\ell_p = \text{malicious} \quad \text{if} \quad \exists q \in p : \ell_q = \text{malicious}.
\]
This principle ensures that any detected malicious component invalidates the package’s overall trustworthiness.
The detection system produces both the binary verdict and a structured report that identifies the responsible slices, their matched historical examples, and the expert reasoning principles supporting the decision.

\section{EVALUATION}
\label{sec:experiment}

We evaluate our tool to answer the following research questions:

 \begin{itemize}
 \item \textbf{RQ1 Detection Accuracy:} How does \tool{} perform compared to state-of-the-art baseline methods in detecting malicious packages? 
 \item \textbf{RQ2 Obfuscation Robustness:} Can \tool{} maintain detection accuracy when malicious code is obfuscated using transformation techniques? 
  \item \textbf{RQ3 Knowledge Extraction:} What expert knowledge is extracted from threat reports and how does it characterize malicious behavior?
  \item \textbf{RQ4 Knowledge Contribution:} What is the contribution of expert knowledge to detection accuracy?
  
\end{itemize}
% \vspace{-10pt}

\begin{table}[]
\centering
\scriptsize
\caption{Detection Performance Comparison on PyPI}
\label{tab:rq2_pypi}
\begin{tabular}{lcccc}
\toprule
\textbf{Tool} & \textbf{Accuracy (\%)} & \textbf{Precision (\%)} & \textbf{Recall (\%)} & \textbf{F1-Score (\%)} \\
\midrule
Bandit4Mal & \cellcolor{red!20}28.65 & \cellcolor{red!20}34.58 & 47.90 & 40.17 \\
GuardDog & 87.30 & 91.54 & 82.20 & 86.62 \\
OSSGadget & 51.90 & 51.10 & 87.90 & 64.63 \\
PyPIWarehouse & 53.25 & 51.69 & \cellcolor{green!30}99.30 & 67.99 \\
SocketAI & 90.30 & 93.06 & 87.10 & 89.98 \\
SAP-\textsubscript{DT} & 60.90 & 90.07 & \cellcolor{red!20}24.50 & \cellcolor{red!20}38.52 \\
SAP-\textsubscript{RF} & 88.95 & 98.27 & 79.30 & 87.77 \\
SAP-\textsubscript{XGB} & 66.30 & 92.89 & 35.30 & 51.16 \\
MalGuard-\textsubscript{DT} & 90.35 & \cellcolor{green!30}100.00 & 80.70 & 89.32 \\
MalGuard-\textsubscript{MLP} & 94.45 & \cellcolor{yellow!30}99.89 & 89.00 & 94.13 \\
MalGuard-\textsubscript{NB} & 63.35 & 98.19 & 27.20 & 42.60 \\
MalGuard-\textsubscript{RF} & 90.15 & \cellcolor{green!30}100.00 & 80.30 & 89.07 \\
MalGuard-\textsubscript{SVM} & \cellcolor{yellow!30}94.75 & \cellcolor{yellow!30}99.89 & 89.60 & \cellcolor{yellow!30}94.46 \\
MalGuard-\textsubscript{XGB} & 89.65 & \cellcolor{green!30}100.00 & 79.30 & 88.46 \\
\midrule
\textbf{\tool{} (Ours)} & \cellcolor{green!30}\textbf{99.00} & 99.49 & \cellcolor{yellow!30}\textbf{98.50} & \cellcolor{green!30}\textbf{98.99} \\
\bottomrule
\end{tabular}
% \vspace{2mm}
% \begin{minipage}{\linewidth}
% \footnotesize
% \textbf{Note:} \colorbox{green!30}{Green} indicates the highest value in each column, 
% \colorbox{yellow!30}{yellow} indicates the second-highest value, 
% and \colorbox{red!20}{red} indicates the lowest value.
% \end{minipage}
% \vspace{-5mm}
\end{table}

\noindent\textbf{Dataset Construction.}
We construct test datasets for npm and PyPI, ensuring no overlap with packages 
used in knowledge base construction to prevent data leakage. Each dataset 
contains 1,000 malicious packages (from established datasets by Ohm et 
al.~\cite{ohm2020backstabber} for npm and Guo et al.~\cite{guo2023empirical} 
for PyPI) and 1,000 benign packages sampled from the top 5,000 
most downloaded packages on each registry. This sampling strategy ensures 
benign samples represent widely-used production code rather than obscure packages.
To evaluate robustness against obfuscation, we create an additional PyPI test 
set with 1,041 malicious and 1,019 benign packages transformed using Intensio-Obfuscator~\cite{intensio2020}. The obfuscation applies variable renaming, control flow flattening, and string encoding to assess whether \tool{} maintains accuracy 
when syntactic signatures are deliberately obscured. All datasets are shown in Appendix~\ref{tab:dataset}

\noindent\textbf{Baseline Methods.}
We select SOTA detection methods based on four criteria: 
\textit{diversity} (representing different detection paradigms including 
static analysis, dynamic analysis, machine learning, and LLM-based approaches), 
\textit{reproducibility} (publicly available with clear documentation), 
\textit{recency} (actively maintained with recent updates), and \textit{impact} 
(developed by major organizations or widely adopted in practice).
Table~\ref{tab:baselines} summarizes our selected baselines (detailed in 
Appendix). For static analysis, we include OSSGadget (Microsoft)~\cite{ossgadget2021} and GuardDog 
(Datadog)~\cite{guarddog2022}, both supporting PyPI and npm. For machine learning, we evaluate 
SAP~\cite{ladisa2023feasibility} with three variants (i.e., ~\cite{chen2016xgboost}, Random Forest~\cite{breiman2001random}, Decision Tree~\cite{song2015decision}) supporting 
both ecosystems, alongside ecosystem-specific methods: MalGuard~\cite{gao2025malguard} for PyPI and 
MalPacDetector~\cite{wang2025malpacdetector} for npm. For dynamic analysis, we select Packj~\cite{packj2022}, which monitors 
runtime behaviors across both ecosystems. Finally, we include SocketAI~\cite{zahan2025leveraging} as 
the LLM-driven baseline applicable to PyPI and npm.

\noindent\textbf{Experimental Setup.}
In our experiments, we employ OpenAI's GPT-4.1 model for knowledge extraction from threat intelligence reports and the text-embedding-3-large model (3,072 dimensions) for generating code embeddings during RAG knowledge base construction. For online detection, \tool{} uses GPT-4.1-mini to perform reasoning-enhanced analysis with retrieved expert knowledge. For the SocketAI baseline, we replicate its three-stage analysis pipeline (Initial Report Generation, Critical Reports Analysis, and Final Report Synthesis) following the methodology outlined in the original paper. To balance computational cost and performance, we use GPT-4.1-mini as the large model and GPT-4.1-nano as the small model for each analysis stage, applying the prompts provided in the original work. SocketAI analyzes each code file in a package 

sequentially, a package is classified as malicious if any individual file is identified as malicious in the final synthesis stage. All other baseline tools are executed using their default configurations as specified in their respective documentation.

\subsection{RQ1: Detection Accuracy}

\begin{table}[t]
\centering
\scriptsize
\caption{Detection Performance Comparison on NPM}
\label{tab:rq2}
\begin{tabular}{lcccc}
\toprule
\textbf{Tool} & \textbf{Accuracy (\%)} & \textbf{Precision (\%)} & \textbf{Recall (\%)} & \textbf{F1-Score (\%)} \\
\midrule
Genie                        & 78.79 & 99.67                      & 58.37 & 73.62 \\
GuardDog                     & 94.38 & 96.35                      & 92.41 & 94.34 \\
MalPac-\textsubscript{MLP}   & 94.47 & 97.12                      & 91.83 & 94.40 \\
MalPac-\textsubscript{NB}    & 93.34 & 98.48                      & 88.23 & 93.07 \\
MalPac-\textsubscript{SVM}   & \cellcolor{yellow!30}95.31 & 98.95                      & 91.73 & \cellcolor{yellow!30}95.20 \\
OSSGadget                    & \cellcolor{red!20}56.39 & \cellcolor{red!20}54.23 & 89.79 & \cellcolor{red!20}67.62 \\
Packj\_static                & 57.38 & 54.39 & \cellcolor{green!30}98.74 & 70.15 \\
Packj\_trace                 & 66.21 & 60.46                      & 96.40 & 74.32 \\
SAP-\textsubscript{DT}       & 80.47 & \cellcolor{green!30}100.00 & 67.40 & 80.53 \\
SAP-\textsubscript{RF}       & 80.79 & \cellcolor{green!30}100.00 & 67.94 & 80.91 \\
SAP-\textsubscript{XGB}      & 89.61 & 96.39                      & 85.88 & 90.83 \\
SocketAI                     & 70.94 & 98.03 & \cellcolor{red!20}43.58 & 60.34 \\
\midrule
\textbf{\tool{} (Ours)}      & \cellcolor{green!30}\textbf{98.72} & \cellcolor{yellow!30}\textbf{99.80} & \cellcolor{yellow!30}\textbf{97.67} & \cellcolor{green!30}\textbf{98.72} \\
\bottomrule
\end{tabular}
\end{table}

\noindent\textbf{Overall results.}
Tables~\ref{tab:rq2_pypi} and~\ref{tab:rq2} present detection performance 
on PyPI and npm datasets. \tool{} achieves 99\% accuracy on PyPI with 
99.49\% precision and 98.50\% recall, outperforming the best baseline 
MalGuard-SVM by 4.25 percentage points. On npm, \tool{} achieves 98.72\% 
accuracy with 99.80\% precision and 97.67\% recall, surpassing OSSGadget 
by 31.1 percentage points. In contrast, baselines exhibit significant 
limitations: rule-based tools suffer from extreme false positive rates 
(Bandit4Mal: 90.60\% on PyPI; OSSGadget: 77.98\% on npm), machine learning 
methods show high false negatives (SAP-DT: 75.50\% missed malware on PyPI), 
and SocketAI demonstrates ecosystem-specific inconsistency (19.36 percentage 
point gap between PyPI at 90.30\% and npm at 70.94\%).

\noindent\textbf{Why baselines fail.}
Rule-based detectors match syntax without semantic context, misclassifying benign packages containing documentation URLs or common commands.
ML methods fail due to feature engineering limitations, as sophisticated malware evades detection through semantic variable names and API indirection.
SocketAI struggles with JavaScript-specific techniques and cross-file attack chains, causing ecosystem-specific inconsistency.
Detailed analysis is provided in Appendix~\ref{sec:baseline_analysis}.

\noindent\textbf{Why \tool{} succeeds.}
\tool{} maintains consistent accuracy through two key mechanisms. First, 
behavior-weighted retrieval combines code and behavior similarity, ensuring 
that even when malware uses API indirection like \texttt{getattr(\_\_builtins\_\_, 'exec')}, 
the behavior summary captures "dynamic function call to execute arbitrary code" 
and matches documented indirect execution patterns. Second, sensitive code 
slicing extracts complete execution chains including cross-file dependencies, 
generating comprehensive behavior descriptions such as "downloads content from 
remote server during install and executes via subprocess."

Retrieved expert reasoning provides critical context. For instance, the domain 
\texttt{hits.dwyl.com} becomes identifiable as malicious when matched against 
documented typosquatting campaigns rather than evaluated in isolation. 
Environment variable patterns distinguish legitimate CI checks (\texttt{CI}, 
\texttt{TRAVIS}) from sandbox evasion (\texttt{VIRTUAL\_ENV}, \texttt{WINE\_CPU}) 
based on documented anti-analysis techniques in the knowledge base.

\subsection{RQ2: Obfuscation Robustness}

\begin{table}[t]
\centering
\scriptsize
\caption{Detection Performance on Obfuscated PyPI Packages}
\label{tab:rq3_obfuscation}
\begin{tabular}{lcccc}
\toprule
\textbf{Tool} & \textbf{Accuracy (\%)} & \textbf{Precision (\%)} & \textbf{Recall (\%)} & \textbf{F1-Score (\%)} \\
\midrule
Bandit4Mal & 54.76 & 54.40 & 64.75 & 59.12 \\
GuardDog & \cellcolor{yellow!30}88.11 & 97.16 & 78.77 & \cellcolor{yellow!30}87.00 \\
OSSGadget & \cellcolor{red!20}50.83 & \cellcolor{red!20}50.80 & 85.30 & 63.68 \\
PyPIWarehouse & 63.64 & 58.45 & \cellcolor{green!30}97.02 & 72.95 \\
SocketAI & 84.13 & 82.28 & \cellcolor{yellow!30}87.42 & 84.77 \\
SAP-\textsubscript{DT} & 52.86 & 56.58 & \cellcolor{red!20}28.91 & \cellcolor{red!20}38.27 \\
SAP-\textsubscript{RF} & 65.78 & 76.25 & 46.88 & 58.06 \\
SAP-\textsubscript{XGB} & 60.53 & 70.96 & 37.08 & 48.71 \\
MalGuard-\textsubscript{DT} & 81.55 & 99.40 & 63.88 & 77.78 \\
MalGuard-\textsubscript{MLP} & 86.84 & \cellcolor{yellow!30}99.74 & 74.16 & 85.07 \\
MalGuard-\textsubscript{NB} & 60.24 & 99.55 & 21.42 & 35.26 \\
MalGuard-\textsubscript{RF} & 81.26 & \cellcolor{green!30}100.00 & 62.92 & 77.24 \\
MalGuard-\textsubscript{SVM} & 85.44 & 99.47 & 71.57 & 83.24 \\
MalGuard-\textsubscript{XGB} & 79.66 & 99.68 & 59.94 & 74.87 \\
\midrule
\textbf{\tool{} (Ours)} & \cellcolor{green!30}\textbf{96.50} & \textbf{95.84} & \cellcolor{green!30}\textbf{97.31} & \cellcolor{green!30}\textbf{96.57} \\
\bottomrule
\end{tabular}
\end{table}

\noindent\textbf{Overall results.}
Table~\ref{tab:rq3_obfuscation} presents detection performance on obfuscated 
PyPI packages. \tool{} achieves 96.50\% accuracy with only 2.5 percentage 
point degradation from the original test set, substantially outperforming 
all baselines. Machine learning methods suffer severe drops: SAP-RF falls 
23.17 points and MalGuard variants lose 7-15 points. SocketAI degrades 6.17 
points. Rule-based tools show mixed results: GuardDog remains stable, while 
Bandit4Mal's accuracy increases 26.11 points but only because false positives 
explode from 11.44\% to 55.45\%.

\noindent\textbf{Analysis.}
The obfuscation results reinforce findings from RQ1. Machine learning methods 
fail because Intensio Obfuscator randomizes identifiers to 32-character strings, 
dramatically shifting feature vectors (e.g., identifier entropy jumps from 2.8 
to 4.9) outside learned boundaries despite unchanged malicious APIs. SocketAI 
degrades because randomized names (\texttt{execute\_backdoor} → 
\texttt{a8f3d9e2...}) remove semantic cues, lowering confidence scores and 
causing borderline misclassifications.

\tool{} maintains robustness through behavior abstraction that remains stable 
regardless of identifier names. When generating behavior descriptions, the LLM 
focuses on operations rather than variable names. For example, 
\texttt{os.system("wget " + url + " \&\& bash payload.sh")} produces 
"constructs shell command using wget to download from URL and executes via 
bash"—a description that remains nearly identical whether the variable is 
\texttt{target\_url} or \texttt{b2c3d4e5f6a7...}, achieving 0.93 behavior 
similarity despite only 0.42 code similarity. Retrieved expert reasoning 
explains why this behavior indicates compromise independent of naming choices.

\subsection{RQ3: Knowledge Extraction}

Threat intelligence reports contain three types of actionable knowledge: 
malicious code snippets demonstrating attack implementations, behavioral 
descriptions explaining operational impact, and expert reasoning chains 
articulating security violations. Appendix~\ref{tab:steganography-example} 
illustrates a representative example\footnote{\url{https://blog.phylum.io/fake-aws-packages-ship-command-and-control-malware-in-jpeg-files/}}: an npm package extracts ASCII from 
\texttt{logo.jpg} during postinstall and executes it via Function constructor 
to establish C2 communication. Expert reasoning identifies that AWS S3 libraries 
should parse image formats but \textbf{never execute bytes as code}, and 
template engines may use Function() but \textbf{never on binary file data}.

\noindent\textbf{High-level analytical patterns.}
We analyze how experts structure their overall analysis by examining 7,889 
reasoning chains. Experts employ four analytical strategies: \textit{Functional 
violation reasoning} (4,579 cases, 58\%) identifies behavior contradicting 
package purpose, such as data visualization libraries spawning reverse shells. 
\textit{Contextual boundary violation} (1,814 cases, 23\%) flags operations 
in inappropriate execution contexts, such as installation scripts downloading 
executables. \textit{Privilege abuse reasoning} (946 cases, 12\%) identifies 
unauthorized actions using elevated permissions, such as \texttt{setup.py} 
accessing sensitive files. \textit{Temporal anomaly reasoning} (550 cases, 7\%) 
detects actions at unexpected lifecycle stages, such as credential exfiltration 
during import.

\noindent\textbf{Specific security violation patterns.}
We further analyze 14,076 violated expectation statements, which cite five specific violation types: execution context violations (41.5\%), functional boundary violations (25.0\%), permission abuse violations (18.0\%), data flow violations (10.0\%), and isolation violations (5.5\%). Detailed definitions and examples are provided in Appendix~\ref{sec:violation_patterns}.

% \noindent\textbf{Specific security violation patterns.}
% We then examine the fine-grained security principles that experts invoke, 
% analyzing 14,076 violated expectation statements. These statements cite five 
% specific violation types: \textit{Execution context violations} (5,847, 41.5\%) 
% occur when operations are inappropriate for their execution phase, such as 
% installation scripts accessing credentials when they should only configure 
% local environments. \textit{Functional boundary violations} (3,521, 25.0\%) 
% identify behavior contradicting package purpose, such as data libraries 
% enumerating browser cookies. \textit{Permission abuse violations} (2,536, 18.0\%) 
% flag exploitation of elevated privileges, such as \texttt{setup.py} reading 
% SSH keys when it should only write package files. \textit{Data flow violations} 
% (1,408, 10.0\%) detect unauthorized information transmission, such as sending 
% environment variables to external domains. \textit{Isolation violations} 
% (764, 5.5\%) identify breaches of package manager boundaries, such as modifying 
% adjacent dependencies.

\noindent\textbf{Validation.}
Two security researchers independently reviewed 300 randomly sampled entries. For each entry, annotators compared the extracted knowledge against the original threat report, evaluating whether reasoning chains preserved causal relationships, behavioral descriptions matched actual code functionality, and boundary distinctions were technically valid. Annotators achieved 95.7\% inter-rater agreement on whether extractions faithfully represented original reports. Among agreed cases, 92.3\% of violated expectation 
statements correctly captured identified security principles, and 89.6\% of 
boundary distinctions accurately reflected contextual differences separating 
malicious from legitimate behavior. The primary error mode was incomplete 
context (18 cases), where extractions omitted prerequisite information. This validation confirms our knowledge base captures expert analytical frameworks for retrieval-augmented detection.

\subsection{RQ4: Knowledge Contribution}

% \begin{figure}[]
%     \centering
%     \subfigure[PyPI Normal]{
%         \includegraphics[width=0.10\textwidth]{images/pypi_normal.png}
%         \label{fig:pypi_normal}
%     }
%     \hfill
%     \subfigure[PyPI Obfuscated]{
%         \includegraphics[width=0.10\textwidth]{images/pypi_obfuscated.png}
%         \label{fig:pypi_obfuscated}
%     }
    
%     \subfigure[NPM Ecosystem]{
%         \includegraphics[width=0.10\textwidth]{images/npm_ecosystem.png}
%         \label{fig:npm}
%     }
    
%     \caption{Package Ecosystem Comparison}
%     \label{fig:ecosystems}
% \end{figure}

% \begin{figure}[]
%     \centering
%     \subfigure[PyPI Normal]{
%         \includegraphics[width=0.14\textwidth]{images/pypi_normal.png}
%         \label{fig:pypi_normal}
%     }
%     \hfill
%     \subfigure[PyPI Obfuscated]{
%         \includegraphics[width=0.14\textwidth]{images/pypi_obfuscated.png}
%         \label{fig:pypi_obfuscated}
%     }
%     \hfill
%     \subfigure[NPM Ecosystem]{
%         \includegraphics[width=0.14\textwidth]{images/npm_ecosystem.png}
%         \label{fig:npm}
%     }
%     \caption{Package Ecosystem Comparison}
%     \label{fig:ecosystems}
% \end{figure}

\begin{figure}[t]
    \centering
    \includegraphics[width=1\linewidth]{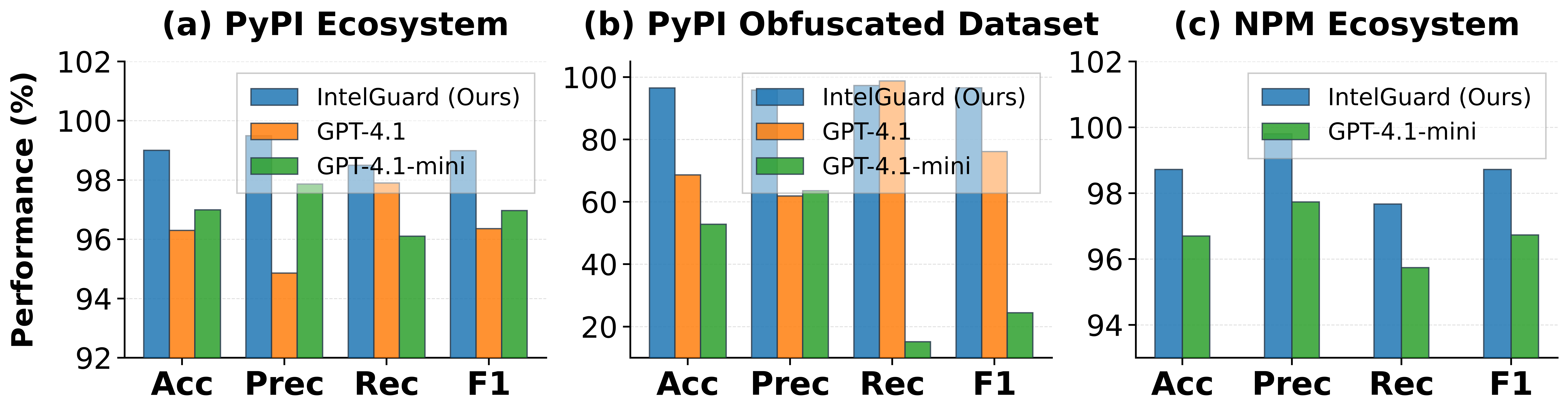}
    \caption{Package Ecosystem Comparison}
    \label{fig:ecosystems}
\end{figure}

\noindent\textbf{Overall impact.}
Figure~\ref{fig:ecosystems} compares \tool{} against direct LLM analysis 
without retrieved knowledge. On standard PyPI evaluation, \tool{} achieves 
99.00\% accuracy with 0.50\% FPR versus GPT-4o-mini's 96.99\% accuracy and 
2.11\% FPR. The gap widens on obfuscated code: \tool{} maintains 
96.50\% accuracy while GPT-4o-mini collapses to 52.82\% with 84.86\% FNR. 
GPT-4o exhibits opposite behavior, flagging most obfuscated benign code as 
malicious (62.12\% FPR). On npm, \tool{} reduces false negatives from 44 to 
24 cases while maintaining lower FPR (0.20\% vs 2.30\%). Most critically, 
retrieved knowledge reduces obfuscated PyPI false negatives from 880 to 28 
cases, a 30x improvement demonstrating that expert reasoning is essential for 
handling code transformations.

\noindent\textbf{Why direct LLM analysis fails.}
Direct LLM analysis lacks two critical capabilities. First, without historical 
attack context, it cannot disambiguate suspicious patterns. When package 
\texttt{anoularjs} performs HTTPS GET to \texttt{hits.dwyl.com} without 
transmitting sensitive data, GPT-4o-mini classifies it as benign because the 
code appears functionally innocuous. \tool{} correctly identifies it as 
malicious by matching against documented typosquatting campaigns from 
\texttt{angulerjs} where \texttt{hits.dwyl.com} served as C2 infrastructure.
Second, without expert guidance on what to examine, LLMs misinterpret obfuscation. 
When encountering randomized identifiers like \texttt{MOEwyUHQwEWWspVGbKLOkTFWMCAaoGUu}, 
GPT models treat obfuscation itself as malicious, misclassifying benign packages. 
Conversely, sophisticated malware using strategic obfuscation causes 84.86\% 
false negatives because the model cannot determine which properties remain 
diagnostic under transformation. Expert knowledge encodes analytical priorities: 
examine API sequences and string literals rather than identifier names, compare 
environment variable usage patterns rather than syntax, and match behavior 
summaries against documented attacks rather than evaluating surface aesthetics.

\noindent\textbf{How retrieved knowledge enables accuracy.}
Retrieved expert reasoning provides contextual disambiguation and analytical focus. 
It supplies historical context for ambiguous indicators and specifies diagnostic properties that remain stable under obfuscation. Detailed examples are provided in Appendix~\ref{sec:knowledge_examples}.

% Retrieved expert reasoning provides contextual disambiguation and analytical 
% focus. For ambiguous indicators, it supplies historical context: \texttt{hits.dwyl.} \texttt{com} 
% becomes diagnostic when matched against typosquatting campaigns rather than 
% evaluated in isolation. For obfuscated code, it specifies what to examine: 
% expert reasoning explains why legitimate CI checks query \texttt{CI} or 
% \texttt{TRAVIS} but sandbox evasion targets \texttt{VIRTUAL\_ENV} or 
% \texttt{WINE\_CPU}, and why legitimate checks log values but evasion code 
% immediately exits. This guidance allows behavior abstraction to work correctly—
% recognizing functionally equivalent operations like \texttt{os.environ.get('VIRTUAL\_ENV')} 
% regardless of variable names, then matching them against anti-sandbox patterns 
% documented in historical attacks.

% \input{body/discussion}
\section{Limitation}
\label{sec:limiation}

% Our work has several limitations.
% First, the knowledge base validation relies primarily on LLM self-assessment, with only 300 entries (approximately 7\%) verified through expert annotation. While LLM-expert correlation (Pearson $r$=0.91) suggests reasonable reliability, systematic bias may exist in unvalidated entries.
% Second, our OCR pipeline achieves 87\% accuracy on code screenshots; recognition errors in complex layouts may propagate to the knowledge base despite subsequent LLM and expert validation.
% Third, although sensitive API slicing reduces attack surface, adversarial prompts targeting LLM-based detection remain a potential threat. Current deployment has not observed such attacks, but integrating dedicated LLM defense mechanisms would strengthen robustness.
% Finally, our evaluation focuses on PyPI and npm ecosystems; generalization to other package registries (e.g., RubyGems, Maven) requires additional validation with ecosystem-specific threat reports.

\noindent\textbf{Real-World Deployment.} We deployed \tool{} to monitor PyPI.org from July 8 to October 1, 2025, analyzing packages at \$0.0016 per package in approximately 5.3 seconds each. We identified 54 previously unreported malicious packages; 24 were officially confirmed by PyPI administrators, while the remaining 30 had been independently removed before our submission (verified as malicious through OSV records). This deployment demonstrates practical viability for real-time ecosystem monitoring.

\noindent\textbf{Limitations.} Our work has several limitations.
First, the knowledge base validation relies primarily on LLM self-assessment, with only 300 entries verified through expert annotation. While LLM-expert correlation (Pearson $r$=0.91) suggests reasonable reliability, systematic bias may exist in unvalidated entries.
Second, our OCR pipeline achieves 87\% accuracy on code screenshots; recognition errors in complex layouts may propagate to the knowledge base despite subsequent validation.
Third, although sensitive API slicing reduces attack surface, adversarial prompts targeting LLM-based detection remain a potential threat. 
% Finally, our evaluation focuses on PyPI and npm; generalization to other registries (e.g., RubyGems, Maven) requires additional validation.
\section{Conclusion}
\label{sec:conclusion}

% We present a knowledge-driven malware detection approach that extracts expert reasoning from threat intelligence reports and applies it through RAG. From 8,024 security reports, we construct a knowledge base capturing behavioral semantics and security violation principles. The system retrieves similar historical attacks and applies expert reasoning to determine malicious intent. Evaluation on 4,087 packages demonstrates 99\% accuracy with 0.50\% false positives, outperforming baselines. On obfuscated code, the approach maintains 96.50\% accuracy while direct LLM analysis dropped to 52.82\%, demonstrating robustness through behavior abstraction and expert knowledge. Between July and September 2025, we deployed the system to detect malicious packages on PyPI.org, identifying 54 new malicious packages and receiving 24 official thanks letters from the administrators.

We present \tool{}, a knowledge-driven framework that extracts expert reasoning from 8,024 threat intelligence reports and applies it through retrieval-augmented generation for malicious package detection. Evaluation on 4,087 packages demonstrates 99\% accuracy with 0.50\% false positives, outperforming baselines. On obfuscated code, \tool{} maintains 96.50\% accuracy while direct LLM analysis achieves only 52.82\%. 
% Deployed on PyPI.org from July to September 2025, the system identified 54 previously unreported malicious packages, with 24 confirmed by the PyPI security team.

% \clearpage
% \balance

\begin{acks}

This work was supported by the National Key Research and Development Program of China (Grant No. 2024YFF0908000). This research is supported by the National Research Foundation, Singapore, and DSO National Laboratories under the AI Singapore Programme (AISG Award No: AISG4-GC-2023-008-1B); by the National Research Foundation Singapore and the Cyber Security Agency under the National Cybersecurity R\&D Programme (NCRP25-P04-TAICeN); and by the Prime Minister’s Office, Singapore under the Campus for Research Excellence and Technological Enterprise (CREATE) Programme.
Any opinions, findings and conclusions, or recommendations expressed in these materials are those of the author(s) and do not reflect the views of the National Research Foundation, Singapore, Cyber Security Agency of Singapore, Singapore.

\end{acks}

\bibliographystyle{ACM-Reference-Format}
\bibliography{acmart}

@article{guo2024packageintel,
  title={PackageIntel: Leveraging Large Language Models for Automated Intelligence Extraction in Package Ecosystems},
  author={Guo, Wenbo and Liu, Chengwei and Wang, Limin and Wu, Jiahui and Xu, Zhengzi and Huang, Cheng and Fang, Yong and Liu, Yang},
  journal={arXiv preprint arXiv:2409.15049},
  year={2024}
}

@article{sarhan2021open,
  title={Open-cykg: An open cyber threat intelligence knowledge graph},
  author={Sarhan, Injy and Spruit, Marco},
  journal={Knowledge-based systems},
  volume={233},
  pages={107524},
  year={2021},
  publisher={Elsevier}
}

@inproceedings{huang2024ctikg,
  title={Ctikg: Llm-powered knowledge graph construction from cyber threat intelligence},
  author={Huang, Liangyi and Xiao, Xusheng},
  booktitle={First Conference on Language Modeling},
  year={2024}
}

@article{hu2024llm,
  title={Llm-tikg: Threat intelligence knowledge graph construction utilizing large language model},
  author={Hu, Yuelin and Zou, Futai and Han, Jiajia and Sun, Xin and Wang, Yilei},
  journal={Computers \& Security},
  volume={145},
  pages={103999},
  year={2024},
  publisher={Elsevier}
}

@article{siracusano2023time,
  title={Time for action: Automated analysis of cyber threat intelligence in the wild},
  author={Siracusano, Giuseppe and Sanvito, Davide and Gonzalez, Roberto and Srinivasan, Manikantan and Kamatchi, Sivakaman and Takahashi, Wataru and Kawakita, Masaru and Kakumaru, Takahiro and Bifulco, Roberto},
  journal={arXiv preprint arXiv:2307.10214},
  year={2023}
}

@inproceedings{guo2023empirical,
  title={An empirical study of malicious code in pypi ecosystem},
  author={Guo, Wenbo and Xu, Zhengzi and Liu, Chengwei and Huang, Cheng and Fang, Yong and Liu, Yang},
  booktitle={2023 38th IEEE/ACM International Conference on Automated Software Engineering (ASE)},
  pages={166--177},
  year={2023},
  organization={IEEE}
}

@inproceedings{huang2024spiderscan,
  title={SpiderScan: Practical detection of malicious NPM packages based on graph-based behavior modeling and matching},
  author={Huang, Yiheng and Wang, Ruisi and Zheng, Wen and Zhou, Zhuotong and Wu, Susheng and Ke, Shulin and Chen, Bihuan and Gao, Shan and Peng, Xin},
  booktitle={Proceedings of the 39th IEEE/ACM International Conference on Automated Software Engineering},
  pages={1146--1158},
  year={2024}
}

@inproceedings{halder2024malicious,
  title={Malicious package detection using metadata information},
  author={Halder, Sajal and Bewong, Michael and Mahboubi, Arash and Jiang, Yinhao and Islam, Md Rafiqul and Islam, Md Zahid and Ip, Ryan HL and Ahmed, Muhammad Ejaz and Ramachandran, Gowri Sankar and Ali Babar, Muhammad},
  booktitle={Proceedings of the ACM Web Conference 2024},
  pages={1779--1789},
  year={2024}
}

@inproceedings{sejfia2022practical,
  title={Practical automated detection of malicious npm packages},
  author={Sejfia, Adriana and Sch{\"a}fer, Max},
  booktitle={Proceedings of the 44th international conference on software engineering},
  pages={1681--1692},
  year={2022}
}

@inproceedings{ferreira2021containing,
  title={Containing malicious package updates in npm with a lightweight permission system},
  author={Ferreira, Gabriel and Jia, Limin and Sunshine, Joshua and K{\"a}stner, Christian},
  booktitle={2021 IEEE/ACM 43rd International Conference on Software Engineering (ICSE)},
  pages={1334--1346},
  year={2021},
  organization={IEEE}
}

@inproceedings{vu2023bad,
  title={Bad snakes: Understanding and improving python package index malware scanning},
  author={Vu, Duc-Ly and Newman, Zachary and Meyers, John Speed},
  booktitle={2023 IEEE/ACM 45th International Conference on Software Engineering (ICSE)},
  pages={499--511},
  year={2023},
  organization={IEEE}
}

@inproceedings{scalco2022feasibility,
  title={On the feasibility of detecting injections in malicious npm packages},
  author={Scalco, Simone and Paramitha, Ranindya and Vu, Duc-Ly and Massacci, Fabio},
  booktitle={Proceedings of the 17th international conference on availability, reliability and security},
  pages={1--8},
  year={2022}
}

@inproceedings{zheng2024towards,
  title={Towards robust detection of open source software supply chain poisoning attacks in industry environments},
  author={Zheng, Xinyi and Wei, Chen and Wang, Shenao and Zhao, Yanjie and Gao, Peiming and Zhang, Yuanchao and Wang, Kailong and Wang, Haoyu},
  booktitle={Proceedings of the 39th IEEE/ACM international conference on automated software engineering},
  pages={1990--2001},
  year={2024}
}

@inproceedings{shariffdeen2025detecting,
  title={Detecting Python Malware in the Software Supply Chain with Program Analysis},
  author={Shariffdeen, Ridwan and Hassanshahi, Behnaz and Mirchev, Martin and El Husseini, Ali and Roychoudhury, Abhik},
  booktitle={2025 IEEE/ACM 47th International Conference on Software Engineering: Software Engineering in Practice (ICSE-SEIP)},
  pages={203--214},
  year={2025},
  organization={IEEE}
}

@inproceedings{gu2023investigating,
  title={Investigating package related security threats in software registries},
  author={Gu, Yacong and Ying, Lingyun and Pu, Yingyuan and Hu, Xiao and Chai, Huajun and Wang, Ruimin and Gao, Xing and Duan, Haixin},
  booktitle={2023 IEEE Symposium on Security and Privacy (SP)},
  pages={1578--1595},
  year={2023},
  organization={IEEE}
}

@inproceedings{huang2024donapi,
  title={$\{$DONAPI$\}$: Malicious $\{$NPM$\}$ Packages Detector using Behavior Sequence Knowledge Mapping},
  author={Huang, Cheng and Wang, Nannan and Wang, Ziyan and Sun, Siqi and Li, Lingzi and Chen, Junren and Zhao, Qianchong and Han, Jiaxuan and Yang, Zhen and Shi, Lei},
  booktitle={33rd USENIX Security Symposium (USENIX Security 24)},
  pages={3765--3782},
  year={2024}
}

@inproceedings{liang2023needle,
  title={A needle is an outlier in a haystack: hunting malicious pypi packages with code clustering},
  author={Liang, Wentao and Ling, Xiang and Wu, Jingzheng and Luo, Tianyue and Wu, Yanjun},
  booktitle={2023 38th IEEE/ACM International Conference on Automated Software Engineering (ASE)},
  pages={307--318},
  year={2023},
  organization={IEEE}
}

@inproceedings{sun20241+,
  title={1+ 1> 2: Integrating deep code behaviors with metadata features for malicious PyPI package detection},
  author={Sun, Xiaobing and Gao, Xingan and Cao, Sicong and Bo, Lili and Wu, Xiaoxue and Huang, Kaifeng},
  booktitle={Proceedings of the 39th IEEE/ACM international conference on automated software engineering},
  pages={1159--1170},
  year={2024}
}

@article{zhang2025killing,
  title={Killing two birds with one stone: Malicious package detection in NPM and PyPI using a single model of malicious behavior sequence},
  author={Zhang, Junan and Huang, Kaifeng and Huang, Yiheng and Chen, Bihuan and Wang, Ruisi and Wang, Chong and Peng, Xin},
  journal={ACM Transactions on Software Engineering and Methodology},
  volume={34},
  number={4},
  pages={1--28},
  year={2025},
  publisher={ACM New York, NY}
}

@inproceedings{li2023malwukong,
  title={MalWuKong: Towards fast, accurate, and multilingual detection of malicious code poisoning in OSS supply chains},
  author={Li, Ningke and Wang, Shenao and Feng, Mingxi and Wang, Kailong and Wang, Meizhen and Wang, Haoyu},
  booktitle={2023 38th IEEE/ACM International Conference on Automated Software Engineering (ASE)},
  pages={1993--2005},
  year={2023},
  organization={IEEE}
}

@inproceedings{ladisa2023feasibility,
  title={On the feasibility of cross-language detection of malicious packages in npm and pypi},
  author={Ladisa, Piergiorgio and Ponta, Serena Elisa and Ronzoni, Nicola and Martinez, Matias and Barais, Olivier},
  booktitle={Proceedings of the 39th annual computer security applications conference},
  pages={71--82},
  year={2023}
}

@inproceedings{yu2024maltracker,
  title={Maltracker: A fine-grained npm malware tracker copiloted by llm-enhanced dataset},
  author={Yu, Zeliang and Wen, Ming and Guo, Xiaochen and Jin, Hai},
  booktitle={Proceedings of the 33rd ACM SIGSOFT International Symposium on Software Testing and Analysis},
  pages={1759--1771},
  year={2024}
}

@inproceedings{gobbi2024genie,
  title={GENIE: guarding the npm ecosystem with semantic malware detection},
  author={Gobbi, Mat{\'\i}as F and Kinder, Johannes},
  booktitle={2024 IEEE Secure Development Conference (SecDev)},
  pages={117--128},
  year={2024},
  organization={IEEE}
}

@inproceedings{zahan2022weak,
  title={What are weak links in the npm supply chain?},
  author={Zahan, Nusrat and Zimmermann, Thomas and Godefroid, Patrice and Murphy, Brendan and Maddila, Chandra and Williams, Laurie},
  booktitle={Proceedings of the 44th International Conference on Software Engineering: Software Engineering in Practice},
  pages={331--340},
  year={2022}
}

@inproceedings{zahan2025leveraging,
  title={Leveraging Large Language Models to Detect npm Malicious Packages},
  author={Zahan, Nusrat and Burckhardt, Philipp and Lysenko, Mikola and Aboukhadijeh, Feross and Williams, Laurie},
  booktitle={2025 IEEE/ACM 47th International Conference on Software Engineering (ICSE)},
  pages={683--683},
  year={2025},
  organization={IEEE Computer Society}
}

@article{gao2025malguard,
  title={MalGuard: Towards Real-Time, Accurate, and Actionable Detection of Malicious Packages in PyPI Ecosystem},
  author={Gao, Xingan and Sun, Xiaobing and Cao, Sicong and Huang, Kaifeng and Wu, Di and Liu, Xiaolei and Lin, Xingwei and Xiang, Yang},
  journal={arXiv preprint arXiv:2506.14466},
  year={2025}
}

@article{wang2025malpacdetector,
  title={MalPacDetector: An LLM-based Malicious NPM Package Detector},
  author={Wang, Jian and Li, Zhen and Qu, Jixiang and Zou, Deqing and Xu, Shouhuai and Xu, Ziteng and Wang, Zhenwei and Jin, Hai},
  journal={IEEE Transactions on Information Forensics and Security},
  year={2025},
  publisher={IEEE}
}

@misc{packj2022,
  author = {Ossillate Inc.},
  title = {Packj: Vetting open-source software packages for security risks},
  year = {2024},
  howpublished = {\url{https://github.com/ossillate-inc/packj}},
  note = {Accessed: 2025-09-10}
}

@misc{ossgadget2021,
  author = {{Microsoft}},
  title = {OSSGadget: Collection of tools for analyzing open source packages},
  year = {2025},
  howpublished = {\url{https://github.com/microsoft/OSSGadget}},
  note = {Accessed: 2025-09-10}
}

@misc{guarddog2022,
  author = {{Datadog Security Labs}},
  title = {GuardDog: A CLI tool to identify malicious PyPI and NPM packages},
  year = {2025},
  howpublished = {\url{https://github.com/DataDog/guarddog}},
  note = {Accessed: 2025-09-10}
}

@misc{bandit4mal2023,
  author = {Vu Duc Ly},
  title = {Bandit4Mal: A Python AST-based tool for detecting malicious PyPI packages},
  year = {2023},
  howpublished = {\url{https://github.com/lyvd/bandit4mal}},
  note = {Accessed: 2025-09-10}}

@misc{PyCG2023,
  author = {Vitalis Salis},
  title = {PyCG: Practical Python Call Graph Generation},
  year = {2023},
  howpublished = {\url{https://github.com/vitsalis/PyCG}},
  note = {Accessed: 2025-10-07}
}

@misc{jelly2025,
  author = {{Computer Science Department, Aarhus University}},
  title = {Jelly: A Static Analysis Framework for JavaScript},
  year = {2025},
  howpublished = {\url{https://github.com/cs-au-dk/jelly}},
  note = {Accessed: 2025-09-10}
}

@misc{intensio2020,
  author = {Hnfull},
  title = {Intensio-Obfuscator: Obfuscate a python code},
  year = {2020},
  howpublished = {\url{https://github.com/Hnfull/Intensio-Obfuscator}},
  note = {Accessed: 2025-09-10}
}

@inproceedings{ohm2020backstabber,
  title={Backstabber’s knife collection: A review of open source software supply chain attacks},
  author={Ohm, Marc and Plate, Henrik and Sykosch, Arnold and Meier, Michael},
  booktitle={International Conference on Detection of Intrusions and Malware, and Vulnerability Assessment},
  pages={23--43},
  year={2020},
  organization={Springer}
}

@misc{cisa2025npm,
  author       = {{Cybersecurity and Infrastructure Security Agency (CISA)}},
  title        = {Widespread Supply Chain Compromise Impacting npm Ecosystem},
  howpublished = {\url{https://www.cisa.gov/news-events/alerts/2025/09/23/widespread-supply-chain-compromise-impacting-npm-ecosystem}},
  note         = {Accessed: 2025-10-07},
  year         = {2025},
  month        = {September},
  organization = {U.S. Department of Homeland Security}
}

@misc{fortinet2025malicious,
  author       = {{Fortinet Threat Research}},
  title        = {Malicious Packages Across Open Source Registries},
  howpublished = {\url{https://www.fortinet.com/blog/threat-research/malicious-packages-across-open-source-registries}},
  note         = {Accessed: 2025-10-07},
  year         = {2025},
  organization = {Fortinet, Inc.}
}

@misc{paloalto2025npm,
  author       = {Asaf Henig and Cameron Hyde},
  title        = {Breakdown: Widespread npm Supply Chain Attack Puts Billions of Weekly Downloads at Risk},
  howpublished = {\url{https://www.paloaltonetworks.com/blog/cloud-security/npm-supply-chain-attack/}},
  note         = {Accessed: 2025-10-07},
  year         = {2025},
  month        = {September},
  organization = {Palo Alto Networks}
}

@article{lu2018learning,
  title={Learning under concept drift: A review},
  author={Lu, Jie and Liu, Anjin and Dong, Fan and Gu, Feng and Gama, Joao and Zhang, Guangquan},
  journal={IEEE transactions on knowledge and data engineering},
  volume={31},
  number={12},
  pages={2346--2363},
  year={2018},
  publisher={IEEE}
}

@inproceedings{duan2021measuring,
  title={Towards Measuring Supply Chain Attacks on Package Managers for Interpreted Languages},
  author={Duan, Ruian and Alrawi, Omar and Kasturi, Ranjita Pai and Elder, Ryan and Saltaformaggio, Brendan and Lee, Wenke},
  booktitle = {28th Annual Network and Distributed System Security Symposium, {NDSS}},
  month     = Feb,
  year      = {2021},
  url       = {https://www.ndss-symposium.org/wp-content/uploads/ndss2021_1B-1_23055_paper.pdf}
}

@misc{osv2025,
  title        = {OSV: A Distributed Vulnerability Database for Open Source},
  howpublished = {\url{https://osv.dev/}},
  note         = {Accessed: 2025-10-27},
  year         = {2025},
  author       = {{Open Source Vulnerabilities Project}},
  organization = {Google Open Source Security Team},
}

@inproceedings{yafei2023libscan,
author = {Wu, Yafei and Sun, Cong and Zeng, Dongrui and Tan, Gang and Ma, Siqi and Wang, Peicheng},
title = {LibScan: towards more precise third-party library identification for android applications},
year = {2023},
isbn = {978-1-939133-37-3},
publisher = {USENIX Association},
address = {USA},
booktitle = {Proceedings of the 32nd USENIX Conference on Security Symposium},
articleno = {190},
numpages = {18},
location = {Anaheim, CA, USA},
series = {SEC '23}
}

@article{xu2024intelex,
  title={IntelEX: A LLM-driven Attack-level Threat Intelligence Extraction Framework},
  author={Xu, Ming and Wang, Hongtai and Liu, Jiahao and Lin, Yun and Liu, Chenyang Xu Yingshi and Lim, Hoon Wei and Dong, Jin Song},
  journal={arXiv preprint arXiv:2412.10872},
  year={2024}
}

@misc{tesseract2025,
  title        = {Tesseract Open Source OCR Engine (GitHub Repository)},
  author       = {{tesseract-ocr}},
  howpublished = {\url{https://github.com/tesseract-ocr/tesseract}},
  note         = {Accessed: 2025-10-27},
  year         = {2025},
}

@misc{reversinglabs2025,
  title        = {ReversingLabs – Software Supply Chain Security \& Threat Intelligence},
  howpublished = {\url{https://www.reversinglabs.com/}},
  note         = {Accessed: 2025-10-27},
  author       = {{ReversingLabs}},
  year         = {2025},
}

@misc{kaspersky2025,
  title        = {Kaspersky Lab},
  howpublished = {\url{https://www.kaspersky.com/}},
  note         = {Accessed: 2025-10-27},
  author       = {{Kaspersky Lab}},
  year         = {2025},
}

@misc{trendmicro2025,
  title        = {Trend Micro – Cybersecurity Solutions},
  howpublished = {\url{https://www.trendmicro.com/}},
  note         = {Accessed: 2025-10-27},
  author       = {{Trend Micro}},
  year         = {2025},
}

@article{mcinnes2017hdbscan,
  title={hdbscan: Hierarchical density based clustering.},
  author={McInnes, Leland and Healy, John and Astels, Steve and others},
  journal={J. Open Source Softw.},
  volume={2},
  number={11},
  pages={205},
  year={2017}
}

@inproceedings{zimmermann2019small,
  title={Small world with high risks: A study of security threats in the npm ecosystem},
  author={Zimmermann, Markus and Staicu, Cristian-Alexandru and Tenny, Cam and Pradel, Michael},
  booktitle={28th USENIX Security symposium (USENIX security 19)},
  pages={995--1010},
  year={2019}
}

@inproceedings{wyss2022wolf,
  title={Wolf at the door: Preventing install-time attacks in npm with latch},
  author={Wyss, Elizabeth and Wittman, Alexander and Davidson, Drew and De Carli, Lorenzo},
  booktitle={Proceedings of the 2022 ACM on Asia Conference on Computer and Communications Security},
  pages={1139--1153},
  year={2022}
}

@inproceedings{li2022attackg,
  title={AttacKG: Constructing technique knowledge graph from cyber threat intelligence reports},
  author={Li, Zhenyuan and Zeng, Jun and Chen, Yan and Liang, Zhenkai},
  booktitle={European Symposium on Research in Computer Security},
  pages={589--609},
  year={2022},
  organization={Springer}
}

@article{rani2024ttpxhunter,
  title={Ttpxhunter: Actionable threat intelligence extraction as ttps from finished cyber threat reports},
  author={Rani, Nanda and Saha, Bikash and Maurya, Vikas and Shukla, Sandeep Kumar},
  journal={Digital Threats: Research and Practice},
  volume={5},
  number={4},
  pages={1--19},
  year={2024},
  publisher={ACM New York, NY}
}

@article{breiman2001random,
  title={Random forests},
  author={Breiman, Leo},
  journal={Machine learning},
  volume={45},
  pages={5--32},
  year={2001},
  publisher={Springer}
}

@inproceedings{chen2016xgboost,
  title={Xgboost: A scalable tree boosting system},
  author={Chen, Tianqi and Guestrin, Carlos},
  booktitle={Proceedings of the 22nd acm sigkdd international conference on knowledge discovery and data mining},
  pages={785--794},
  year={2016}
}

@article{song2015decision,
  title={Decision tree methods: applications for classification and prediction},
  author={Song, Yan-Yan and Lu, Ying},
  journal={Shanghai archives of psychiatry},
  year={2015},
  publisher={Shanghai Municipal Bureau of Publishing}
}

% \section{Appendix}
\appendix

\section{Appendix}
\label{sec:appendix}

\subsection{Structured Knowledge Example}
\label{tab:steganography-example-section}

Table~\ref{tab:steganography-example} presents a representative example of structured knowledge extracted from a Phylum threat report\footnote{\url{https://blog.phylum.io/fake-aws-packages-ship-command-and-control-malware-in-jpeg-files/}}. This case demonstrates how our extraction pipeline captures malicious code snippets, behavioral descriptions, and expert reasoning chains that explain why specific patterns indicate compromise.

\begin{table}[!h]
\centering
\caption{Example of Structured Knowledge Extracted from a Threat Intelligence Report}
\label{tab:steganography-example}
\scriptsize
\begin{tabular}{@{}p{0.22\columnwidth}p{0.75\columnwidth}@{}}
\toprule
\textbf{Knowledge Type} & \textbf{Content} \\
\midrule
\textbf{Malicious Code} & 
\begin{lstlisting}[style=tablecode]
function processImage(filePath) {
    const data = fs.readFileSync(filePath);
    let payload = "";
    for (let i = 0; i < data.length; i++) {
        if (data[i] >= 32 && data[i] <= 126) {
            payload += String.fromCharCode(data[i]);
        }
    }
    // Execute extracted code from image
    const func = new Function('https','exec','os', payload);
    func(https, exec, os);
}
processImage('logo.jpg');
\end{lstlisting} \\
\midrule
\textbf{Behavior} & 
During postinstall, reads image file (\texttt{logo.jpg}), extracts ASCII character sequences (bytes 32-126) hidden in image data, and executes the extracted payload using Function constructor, establishing C2 communication with \texttt{85.208.108.29}. \\
\midrule
\textbf{Expert Reasoning} & 
\textit{Why Suspicious}: S3 library should not process/execute image contents. \textit{Violations}: Postinstall extracts code from images (context violation); images used as executable sources (data flow violation); \texttt{.jpg} files hide code to evade analysis (deceptive technique). \textit{Boundary Distinction}: Legitimate image parsing \textbf{never executes bytes as code}; Function() \textbf{never runs on binary file data}. \\
\bottomrule
\end{tabular}
\end{table}
% \footnotetext{\url{https://blog.phylum.io/fake-aws-packages-ship-command-and-control-malware-in-jpeg-files/}}

\subsection{Evaluation Setup}
\label{sec:eval_setup}

Table~\ref{tab:dataset} summarizes the evaluation datasets. We construct test sets for the NPM and PyPI ecosystems, ensuring no overlap with packages used in the knowledge base construction. Additionally, we create an obfuscated PyPI dataset using Intensio-Obfuscator~\cite{intensio2020} to evaluate robustness against code transformations. Table~\ref{tab:baselines} lists the baseline methods spanning rule-based, machine learning, and LLM-based approaches.

\begin{table}[!h]
\centering
\scriptsize
\caption{Summary of Evaluation Datasets}
\label{tab:dataset}
\resizebox{0.75\columnwidth}{!}{ 
\begin{tabular}{lrrr}
\toprule
\textbf{Dataset} & \textbf{Malicious} & \textbf{Benign} & \textbf{Total} \\
\midrule
NPM & 1,028 & 999 & 2,027 \\
PyPI & 1,000 & 1,000 & 2,000 \\
PyPI-Obfuscated & 1,041 & 1,019 & 2,060 \\
\bottomrule
\end{tabular}
}
\end{table}

\begin{table}[!h]
\centering
\scriptsize
\caption{Baseline Methods for Evaluation}
\label{tab:baselines}
\begin{tabular}{llcc}
\toprule
\textbf{Type} & \textbf{Tool} & \textbf{Year} & \textbf{Target} \\
\midrule
\multirow{2}{*}{Static Rule-Based} 
& OSSGadget~\cite{ossgadget2021} & 2021 & NPM, PyPI \\
& GuardDog~\cite{guarddog2022} & 2022 & NPM, PyPI \\
\midrule
Dynamic Rule-Based 
& Packj~\cite{packj2022} & 2022 & NPM, PyPI \\  
\midrule
\multirow{4}{*}{Machine Learning-Based} 
& SAP (XGBoost)~\cite{ladisa2023feasibility} & 2023 & NPM, PyPI \\
& SAP (Random Forest) & 2023 & NPM, PyPI \\
& SAP (Decision Tree) & 2023 & NPM, PyPI \\
& MalGuard~\cite{gao2025malguard} & 2023 & PyPI \\
\midrule
\multirow{2}{*}{LLM-Based} 
& SocketAI~\cite{zahan2025leveraging} & 2024 & NPM, PyPI \\
& MalPacDetector~\cite{wang2025malpacdetector} & 2025 & NPM \\
\bottomrule
\end{tabular}
\end{table}

\subsection{Temporal Validation}
\label{sec:temporal_validation}

To evaluate \tool{}'s generalization capability on previously unseen threats, we conducted a temporal validation experiment using malicious packages published after the knowledge cutoff date of the underlying LLM.
Specifically, we collected 959 malicious PyPI packages published after January 2025, which is beyond GPT-4.1-mini's knowledge cutoff date (June 2024).
This ensures that the LLM has no prior exposure to these packages during its training phase.

Table~\ref{tab:temporal_validation} presents the detection performance on this temporally separated dataset.
\tool{} achieves 98.31\% accuracy and 98.29\% F1-score, demonstrating strong generalization to novel threats that were not present in either the LLM's training data or our knowledge base construction period.
This result confirms that \tool{}'s effectiveness stems from the transferable expert reasoning patterns rather than memorization of specific package signatures.

\begin{table}[!h]
\centering
\scriptsize
\caption{Temporal Validation on Post-2025 PyPI Packages}
\label{tab:temporal_validation}
\begin{tabular}{lcccc}
\toprule
\textbf{Method} & \textbf{Accuracy (\%)} & \textbf{Precision (\%)} & \textbf{Recall (\%)} & \textbf{F1-Score (\%)} \\
\midrule
\tool{} & 98.31 & 97.83 & 98.75 & 98.29 \\
\bottomrule
\end{tabular}
\end{table}

\subsection{Cost and Runtime Analysis}
\label{sec:cost_analysis}

We analyze the computational cost and runtime efficiency of \tool{} to assess its practical viability for real-world deployment.

\noindent\textbf{Knowledge base construction.}
The one-time cost of constructing the knowledge base using GPT-4.1 for processing 8,024 threat intelligence reports and extracting 4,420 validated knowledge entries is \$60.32.
This cost is amortized over all subsequent detections, as the knowledge base requires only incremental updates when new threat reports become available.

\noindent\textbf{Detection cost and runtime.}
Table~\ref{tab:cost_comparison} compares the per-package detection cost and runtime of \tool{} against baseline methods on the PyPI dataset.
\tool{} achieves 7.5$\times$ lower cost and 3.9$\times$ faster runtime compared to SocketAI while delivering 8.7 percentage points higher accuracy.
Compared to rule-based tools like GuardDog, \tool{} incurs modest computational overhead but provides substantially improved detection accuracy (99.00\% vs. 87.30\%).

\begin{table}[!h]
\centering
\scriptsize
\caption{Cost and Runtime Comparison on PyPI Dataset}
\label{tab:cost_comparison}
\begin{tabular}{lcccc}
\toprule
\textbf{Method} & \textbf{Cost/pkg} & \textbf{Time/pkg (s)} & \textbf{Accuracy (\%)} \\
\midrule
\tool{} & \$0.018 & 29.48 & 99.00 \\
SocketAI & \$0.135 & 114.48 & 90.30 \\
GuardDog & free & 23.20 & 87.30 \\
\bottomrule
\end{tabular}
\end{table}

\noindent\textbf{Open-source model alternative.}
To further reduce deployment costs, we evaluate \tool{} with DeepSeek-V3, an open-source LLM that can be self-hosted at zero API cost.
As shown in Table~\ref{tab:deepseek_comparison}, the DeepSeek-V3 variant achieves 98.75\% accuracy on PyPI, comparable to the GPT-4.1-mini-based configuration.
This demonstrates that \tool{}'s knowledge-driven approach is model-agnostic and can leverage cost-effective open-source alternatives without significant performance degradation.

\begin{table}[!h]
\centering
\scriptsize
\caption{Performance with Open-Source LLM (DeepSeek-V3)}
\label{tab:deepseek_comparison}
\begin{tabular}{lcccc}
\toprule
\textbf{Method} & \textbf{Accuracy (\%)} & \textbf{Precision (\%)} & \textbf{Recall (\%)} & \textbf{F1-Score (\%)} \\
\midrule
\tool{} (DeepSeek-V3) & 98.75 & 98.60 & 98.91 & 98.76 \\
\bottomrule
\end{tabular}
\end{table}

\subsection{Detailed Analysis of Baseline Limitations (RQ1)}
\label{sec:baseline_analysis}

\noindent\textbf{Why rule-based tools fail.}
Rule-based detectors match syntax without semantic context. Bandit4Mal flags 
any URL string or file read operation as suspicious, misclassifying 906 of 
1,000 benign PyPI packages containing documentation URLs or configuration 
reads. It cannot distinguish \texttt{config.read()} from malicious operations 
like \texttt{os.system(open('/etc/passwd').read())}. Similarly, OSSGadget's 
LOLBAS rules flag common commands (\texttt{apt-get}, \texttt{curl}) regardless 
of context, treating legitimate installation scripts as backdoors.

\noindent\textbf{Why machine learning fails.}
ML approaches fail due to feature engineering limitations. SAP extracts 
identifier entropy and API call counts, but sophisticated malware evades 
these features through semantic variable names and API indirection 
(\texttt{getattr(\_\_builtins\_\_, 'exec')}), maintaining low entropy while 
avoiding direct suspicious matches. When malicious logic spreads across 
functions with normal statistical profiles, decision trees misclassify based 
on overly simplistic rules.

\noindent\textbf{Why SocketAI shows ecosystem gaps.}
SocketAI's performance gap stems from training data imbalance and limited 
analysis scope. The model handles Python patterns (\texttt{exec(compile(...))}) 
well but struggles with JavaScript techniques 
(\texttt{new Function(atob())}, \texttt{Object.constructor.constructor()}). 
Critically, SocketAI analyzes files independently, missing cross-file attack 
chains where \texttt{package.json} triggers \texttt{install.js} which loads 
\texttt{payload.js}. For base64-encoded payloads, it cannot determine decoded 
content statically, defaulting to benign classification.

\balance

\subsection{Security Violation Pattern Details (RQ3)}
\label{sec:violation_patterns}

We analyze 14,076 violated expectation statements from expert reasoning chains. These statements cite five specific violation types:
\textit{Execution context violations} (5,847, 41.5\%) occur when operations are inappropriate for their execution phase, such as installation scripts accessing credentials when they should only configure local environments. 
\textit{Functional boundary violations} (3,521, 25.0\%) identify behavior contradicting package purpose, such as data libraries enumerating browser cookies. 
\textit{Permission abuse violations} (2,536, 18.0\%) flag exploitation of elevated privileges, such as \texttt{setup.py} reading SSH keys when it should only write package files. 
\textit{Data flow violations} (1,408, 10.0\%) detect unauthorized information transmission, such as sending environment variables to external domains. 
\textit{Isolation violations} (764, 5.5\%) identify breaches of package manager boundaries, such as modifying adjacent dependencies.

\subsection{How Retrieved Knowledge Enables Accuracy (RQ4)}
\label{sec:knowledge_examples}

Retrieved expert reasoning provides contextual disambiguation and analytical 
focus. For ambiguous indicators, it supplies historical context: \texttt{hits.dwyl.} \texttt{com} 
becomes diagnostic when matched against typosquatting campaigns rather than 
evaluated in isolation. For obfuscated code, it specifies what to examine: 
expert reasoning explains why legitimate CI checks query \texttt{CI} or 
\texttt{TRAVIS} but sandbox evasion targets \texttt{VIRTUAL\_ENV} or 
\texttt{WINE\_CPU}, and why legitimate checks log values but evasion code 
immediately exits. This guidance allows behavior abstraction to work correctly—
recognizing functionally equivalent operations like \texttt{os.environ.get('VIRTUAL\_ENV')} 
regardless of variable names, then matching them against anti-sandbox patterns 
documented in historical attacks.

\subsection{Discussion of Knowledge Update}

The knowledge base supports incremental updates from new threat reports. When vendors such as ReversingLabs, Checkmarx, or Phylum publish threat analyses, we apply the extraction pipeline to obtain malicious code snippets, behavioral descriptions, and reasoning chains. New entries integrate into the knowledge base through dual-embedding indexing and become immediately available for retrieval. This enables adaptation to evolving threats without modifying the detection logic. The knowledge base currently covers PyPI and NPM. The behavior-based abstraction operates independently of language syntax, allowing straightforward extension to RubyGems, NuGet, or Maven by processing their threat reports with the same extraction pipeline.

\subsection{Case Study: readosso-0.1.0}

We present one representative case from the 54 detected packages to illustrate \tool{}'s detection capability. In PyPI, each package is uniquely identified by its (name, version) tuple, and removed malicious package names cannot be reused. Therefore, all 54 detected packages are first-time discoveries. Clone detection confirmed that 52 have distinct implementations.

On September 19, 2025, \tool{} identified \texttt{readosso-0.1.0} as malicious. The package embeds a Discord bot backdoor that automatically executes upon import and downloads arbitrary executables from attacker-controlled URLs for execution via \texttt{os.startfile()}.

\begin{lstlisting}[
    language=Python,
    caption={Malicious code found in readosso-0.1.0 package},
    label={lst:malicious-code},
    basicstyle=\ttfamily\small,
    keywordstyle=\color{blue}\bfseries,
    commentstyle=\color{gray}\itshape,
    stringstyle=\color{red},
    numbers=left,
    numberstyle=\tiny\color{gray},
    stepnumber=1,
    numbersep=8pt,
    showstringspaces=false,
    breaklines=true,
    frame=single,
    backgroundcolor=\color{lightgray!10}
]
@bot.command()
async def addprogram(ctx):
    if ctx.guild and ctx.guild.id == GUILD_ID:
        exe_url = msg.content.strip()
        appdata = os.getenv('APPDATA')
        exe_path = os.path.join(appdata, "downloaded_program.exe")
        r = requests.get(exe_url)
        with open(exe_path, "wb") as f:
            f.write(r.content)
        os.startfile(exe_path)

if "readosso" in sys.modules or __name__ == "__main__":
    import threading
    t = threading.Thread(target=run_bot, daemon=True)
    t.start()
\end{lstlisting}

\begin{figure}
\centering
\includegraphics[width=0.4\textwidth]{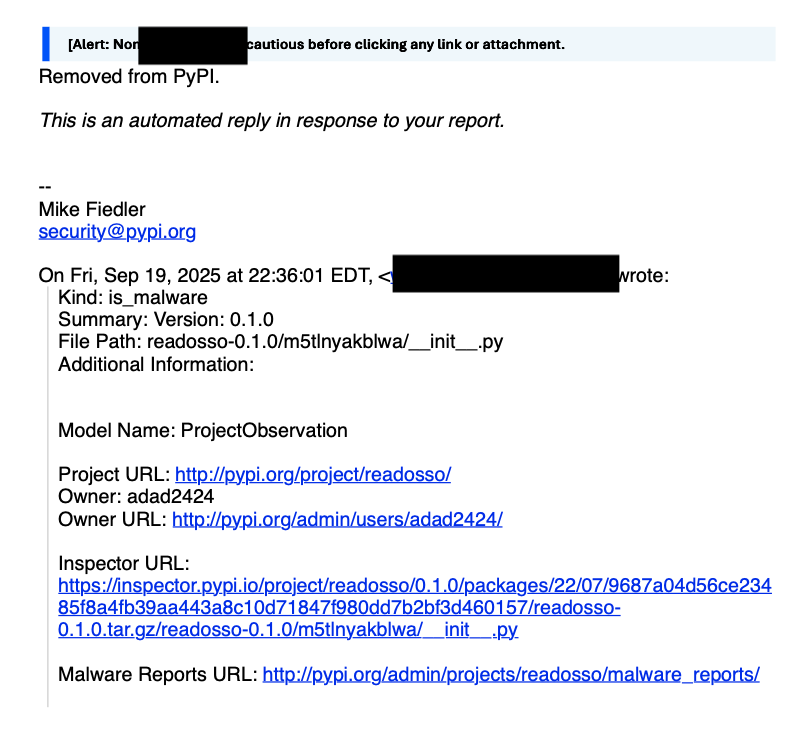}
\caption{PyPI security team confirmation for readosso-0.1.0 removal}
\label{fig:readosso_confirmation}
\end{figure}

\tool{} retrieved similar backdoor patterns from historical packages and applied expert reasoning that installation-time external communication and arbitrary executable download violate functional boundaries. We reported this finding to PyPI administrators on September 19, 2025. On September 22, 2025, the PyPI security team confirmed the package as malicious and removed it in Figure~\ref{fig:readosso_confirmation}.

\end{document}